\documentclass[twoside]{article}

\usepackage{PRIMEarxiv}

\usepackage[utf8]{inputenc} 
\usepackage[T1]{fontenc}    
\usepackage{hyperref}       
\usepackage{url}            
\usepackage{booktabs}       
\usepackage{amsfonts}       
\usepackage{nicefrac}       
\usepackage{lipsum}
\usepackage{fancyhdr}       
\usepackage{graphicx}       
\usepackage{float}
\usepackage{caption}
\usepackage{setspace}
\graphicspath{{Figures/}}     

\title{Automated Unit Test Case Generation: A Systematic Literature Review
}

\author{
  Jason Wang \\
  University of Sydney \\
  Sydney, Australia \\
  \texttt{jwan7009@uni.sydney.edu.au} \\
   \And
  Basem Suleiman \\
  University of New South Wales \\
  Sydney, Australia \\
  \texttt{b.suleiman@unsw.edu.au} \\
    \And
  Muhammad Johan Alibasa \\
  Monash University Indonesia\\
  Tangerang, Indonesia \\
  \texttt{johan.alibasa@monash.edu} \\
}
  
\begin{document}
\maketitle

\begin{abstract}
Software is omnipresent within all factors of society. It is thus important to ensure that software are well tested to mitigate bad user experiences as well as the potential for severe financial and human losses. Software testing is however expensive and absorbs valuable time and resources. As a result, the field of automated software testing has grown of interest to researchers in past decades. In our review of present and past research papers, we have identified an information gap in the areas of improvement for the Genetic Algorithm and Particle Swarm Optimisation. A gap in knowledge in the current challenges that face automated testing has also been identified. We therefore present this systematic literature review in an effort to consolidate existing knowledge in regards to the evolutionary approaches as well as their improvements and resulting limitations. These improvements include hybrid algorithm combinations as well as interoperability with mutation testing and neural networks. We will also explore the main test criterion that are used in these algorithms alongside the challenges currently faced in the field related to readability, mocking and more. 
\end{abstract}

\clearpage

\section{Introduction}

Software is ubiquitous in our world today and is prevalent in all aspects of our daily lives. It is present in all aspects of our lives ranging from applications for heavy machinery and transport systems all the way to the small mobile devices that reside within our pockets that enable us to communicate with one another. McKinsey \& Co have stated that "software is the world" and that there will be a point where "every company is a software company" \cite{Mckinsey}. As a major component of society with endless applications supporting individuals in their daily lives, there can therefore be minor or severe ramifications when software fails. Software development is not without error and bad programming practices will lead to unintended consequences.  Software developers therefore need to test their code and applications with the various testing methodologies and tools available at their disposal.


Software has failed in the past, and it will continue to do so into the future. Software failures can be a simple bug in a mobile or web application resulting in a unsatisfactory user experience. Catastrophic failures that result in the loss of life, assets and/or capital can also arise as a result of software failures. High profile incidents of software failure include the total loss of the European Space Agency's 1996 maiden launch of the Arianne V rocket. Bentley reported that the failure resulted in an uninsured loss of \$500,000,000 and setbacks in development. The failure was due to a "lack of exception handling for a floating-point error when a 64-bit integer was converted into a 16-bit signed integer" causing it to self-destruct and thus highlights the fact that "software testing is one of the most critical and important phases in software testing" \cite{Kosindredcha-TestCase}. Other cases of inadequate software testing include the 2012 Knights Capital Group's "trading glitch" that cost the company of its leadership as well as a loss of \$440,000,000. A poorly tested trading algorithm that was hastily put together resulted in Knights buying large volumes of stocks within the first 30 minutes of trading without the equity to back up their purchases. In additional financial metrics, NIST in 2002 reported that 0.6\% of the US GDP was lost as a result of software failures \cite{Jia-PSO-Test-Generation}. Inadequate software testing therefore leads to software failures that can have a plethora of unintended effects ranging from mild annoyances all the way to dramatic monetary losses and the loss of life. 

The usage of software and the assurance that it runs as designed and intended without failure is therefore a very important aspect that software developers are faced with. InadequRodriguesate testing can have real world impacts as seen in the case studies above and can also be the root causes of problems including: (1) Increased failures due to poor quality; (2) Increased software development costs; (3) Increased time to market due to inefficient testing; and (4) increased market transaction costs, which can impact development budgets \cite{Kosindredcha-TestCase}. Ensuring adequate software testing practices is therefore a very important factor in the software development life-cycle. 


There are currently various forms of testing methodologies a developer can apply to their testing strategy. Testing can be as straightforward as getting users and stakeholders to use the software. This involves users interacting with the software to see if certain features and functions work as intended. This form of testing is known as "Acceptance Testing" and is done to ensure that the software is developed as per the requirements of the customer/stakeholder \cite{Gaur-Testing-Techniques}. This form of testing is generally high level, involves multiple rounds of testing such as an alpha and beta testing phase and is generally not dependent on any additional coding. 

Diving down into the lower level of testing is where "Unit Testing" resides and is where individual functions/modules are tested prior to being integrated with other modules and components. Unit testing is the first level of testing and has the advantages of low cost error finding as the size of the single module is small and decoupled. The small module size also provides an additional benefit in which it can be tested a lot quicker \cite{Gaur-Testing-Techniques}. 
This type of testing allows the developer to tests isolated components in smaller segments and allows them to ensure that those small components are satisfactory in their behaviour prior to further integration. Another form of testing that is now actively growing in usage by developers is "Automatic Testing" which is the main topic of this paper. Automated Testing involves a series of tools and algorithms that allow developers to generate unit test cases automatically for various code-bases without the need for the developer to actively code for them. This approach saves developers time and resources of which can be redirected elsewhere. 

A plethora of research have indicated that the software testing stage of the development life-cycle takes up a significant amount of time and resources. 
Software testing eats into this time to the point where many previous papers state that "software testing often accounts for more than 50\% of the total development process" \cite{Anand-Orchestrated-Survey} \cite{Zhu-Improved-GA-Path}. 
Liu et al. \cite{Liu-RBF-NN} states that the design of test cases "requires a lot of manpower, material resources and time". Manual software testing by developers is therefore a long and laborious process. Anand et al. \cite{Anand-Orchestrated-Survey} also reports that "test case generation is one of the most intellectually demanding tasks and is also one of the most critical challenges to general development as it can have a strong impact on the efficiency and efficacy of the whole testing process".

The goal of this paper is to present a critical analysis and review of automated test case generation. This systematic literature review will build upon previous literature reviews by examining the identified Automatic Unit Test Case Generation (AUTG) algorithms in detail and providing a summary on firstly how the algorithms in question work, their identified flaws and weaknesses when generating test cases, their resultant performance as compared to each other and finally new improvements to those algorithms. Improvements to these algorithms include the fusion of the best components of multiple automated approaches to form an overall better algorithm. Finally, the current main challenges and limitations that plague the field of AUTG will also be explored. 


AUTG as a field is very broad and it would be infeasible for us to present a complete view. The following caveats have therefore been implemented in order to guide and focus the review. The primary goal of this paper is to present a focused view into "White Box Search-Based Software Testing (SBST)" alongside the pertaining meta-heuristic evolutionary algorithms that SBST incorporates and utilises. The paper will therefore present an overview into evolutionary search algorithms such as the Genetic Algorithm as well as Particle Swarm Optimisation. Finally, this paper will also discuss the various improvements that have been made to the two identified algorithms whilst also discussing the limitations that AUTG face in general.

The following research questions will form the basis of the analysis and direction that this paper will undertake:

\begin{itemize}
    \item RQ1 - How do the evolutionary search-based methods of GA and PSO operate? What is their performance as well as the performance of AUTG in general? 
    \item RQ2 - What attempts at improvements have been made in improving GA and PSO?
    \item RQ3 - What are the main challenges and limitations still facing the field of AUTG?
\end{itemize}

\textbf{RQ1} aims to present the "mainly-used" evolutionary search-based AUTG approaches known as the Genetic Algorithm and Particle Swarm Optimisation. We will explore each algorithm in detail and better understand how the methods themselves operate. The additional question of \textbf{RQ1} will also explore the performance of the two approaches alongside performance of AUTG in general. \textbf{RQ2} will explain the various attempts of improvements that have been made by researchers. These improvements include the hybrid combination of GA with another algorithm, its attempts at overcoming the local search optimum solution phenomena and more. Finally, \textbf{RQ3} will explore the the various challenges and limitations that still face AUTG methods today. Some of these issues include the inability to work without type information as well as the inability to work with certain environments such as the web DOM. These results thus interfere and affect the ability for AUTG methods to be utilised more widely in industry. 




As identified earlier, most of the review papers state that more research is required to make the assertion that automated testing is superior. This research gap also includes a lack of comparison and understanding in the various performance benefits and improvements that have been made to the various algorithms. It also includes a major gap in understanding the limitations of these algorithms and what the field of AUTG faces in general. 
We thus aim to review more papers to attempt to make that assertion whilst also exploring other aspects that were not explored earlier. An example of this is to explore the hybrid improvements that can be made to GAs in order to improve its performance and to mitigate their flaws. This concept was briefly explored by Rodrigues et al. \cite{Rodrigues-SysMapping} but not in depth. Some papers have also briefly discussed the problem of GA being stuck in a local optima but we have yet to see a review that looks into this issue in depth. We have also not identified a paper that explores the various problems that are currently being faced by not just GA and PSO but the AUTG field in general. Our paper therefore aims to provide a comprehensive review that addresses these views.



The remainder of this paper will be split across the following sections. Section 2 presents analysis of the related work and discusses research gaps. Section 3 presents the underlying approach that was taken in retrieving all of the studies that were collated for this systematic literature review. Section 4 discusses the various test criterion that automated approaches adapt. Test case criterion is important as it provides a continuous and quantitative measure on how good a particular test case is and their usefulness to the overall test case suite. This will also provide a foundation for the reader prior to our dive into GA and PSO and their accompanied improvements. Section 5 presents an overview of the automated approaches identified and provide an understanding on their internals and how the various algorithms operate to generate test cases. Section 6 discusses the various performance outcomes of the various algorithms and how they fare against each other in terms of performance and usefulness. It also provides analysis of the performance differences that are present between Random Testing and SBST. In addition, this section explores the performance gains that GA and PSO can achieve through the improvements applied such as through Deep Neural Networks and the fusion of other approaches to create hybrid algorithms. A deep dive into the problem of "Genetic Drift" will also be explored. Section 7 presents the current challenges and limitations that plague not just the SBST methods of GA and PSO but automated testing in general. Although very powerful and time saving, automated testing suffers from readability issues as well as its inability to generate test cases for dynamically-typed languages and DOMs that are used by web languages. 


\section{Related Work} \label{chap:relatedwork}

\subsection{Past Literature Reviews}

As a result of our search parameters (Section~\ref{Stage_1_Filtering}), five past papers were identified that made some sort of systematic literature review, mapping or survey into the field of automated testing. The scope of these reviews were either broad and encompassed automated testing as a whole or were niche and specifically focused on certain meta heuristic algorithms such as the Genetic Algorithm (GA) and Particle Swarm Optimisation (PSO). The five papers are listed within Table~\ref{table: previous} and we will now provide a brief summary and commentary as to their contents. 

{
    \renewcommand{\arraystretch}{2}
    \begin{table}[h]
        \centering
        \caption{Previous Systematic Reviews related to Automatic Testing}
        \label{table: previous}
        \begin{tabular}{p{12cm}cc}
            \toprule
            \centering{Title} & Author & Year \\
            \midrule
            Human-based Test Design versus Automated Test Generation: A Literature Review and Meta-Analysis & Kurmaku et al. & 2022 \\
            Using Machine Learning to Generate Test Oracles: A Systematic Literature Review & Fontes and Gay & 2021 \\
            Using Genetic Algorithm in Test Data Generation: A Critical Systematic Mapping & Rodrigues et al. & 2019 \\
            An orchestrated survey of methodologies for automated software test case generation & Anand et al. & 2013 \\
            A Systematic Review of the Application and Empirical Investigation of Search-Based Test Case Generation & Ali et al. & 2010 \\
            \bottomrule
        \end{tabular}
    \end{table}
}

Within their literature review into SBST, Ali et al. \cite{Ali-SysReview} reports that there has been a tremendous amount of research in applying meta-heuristic search algorithms to test case generation. Search-Based Software Testing (SBST) and by extension, the field of Automated Unit Test Case Generation (AUTG), have a very long history in the academic world dating all the way from 1996 to now to which this paper is now aiming to contribute towards. Through its definition, search-based testing is more structured and direct with its generation of test cases through the usage of heuristic algorithms rather than being dependent on a random testing approach that has no search parameters and relies solely on brute force. Testing, as stated previous, is a time and resource expensive expenditure. Ali et al.  \cite{Ali-SysReview} thus presents a comprehensive review into the cost-effective analysis of SBST performed by various other papers as compared to other techniques. The review also analyses whether or not the investigations and findings by other papers were performed correctly. 

The review found that a significant amount of papers related to SBST techniques were focused at the unit-testing level with Genetic Algorithm at the forefront of algorithms in use. The salient observation was that other researchers were not factoring in "random variation" within their test results nor had a baseline comparison point to compare SBST techniques against. This meant that those particular papers were deemed as non-credible and as a result, "the number of papers which contained well-designed and reported empirical studies in the domain of test case generation utilising SBST was very small". Whilst some of the studies selected in the review had credible "effectiveness" results and analysis, the paper overall created an observation where there was limited credible evidence that demonstrated that SBST techniques were even effective as an approach for automated testing over other approaches \cite{Ali-SysReview}. Whilst ultimately not directly relevant with the research goals of this paper, this review provides historical context in the importance of proper reviews. Ali et al. state that it is definitely still the case that SBST techniques are superior to other approaches but ultimately conclude and argue that there isn't enough proper credible research to support those claims and that future papers will need to have better baseline comparisons to properly demonstrate SBST effectiveness against other automated approaches. 

Anand et al. \cite{Anand-Orchestrated-Survey} on the other hand takes a few steps back and provides an orchestrated survey summary of the predominant techniques that are present with the automated testing world. They argue that "the field of software testing is today so vast and specialised that no single author can harness the expertise required for all of the different techniques". Their review contributed towards creating a comprehensive and short overview of each method including random testing and SBST methods. The paper also explores other methods that are out of scope for this paper such as Model-based Testing and Combinatorial Testing. Apart from giving an overview on each automated approach, Anand et al. also discusses performance, possible improvements as well as issues related to each of the above approaches. 

A more recent systematic mapping by Rodrigues et al. \cite{Rodrigues-SysMapping} focuses purely on the usage and application of Genetic Algorithms in test case generation. Their paper also provided a deep analysis into their performance. The authors also comment that their systematic mapping is a lot more specific than the past systematic reviews.
Overall, the review into Genetic Algorithms has found it to be effective as an approach to generate test data successfully due to their "non-deterministic" nature which allows a large solution space to be explored effectively as programs scale upwards and deal with more complex data. GAs also represent an improvement as compared to other "wasteful" approaches such as random testing. Most papers within the mapping have focused on the performance characteristics of the Genetic Algorithm through the adaption of the fitness function. Other characteristics also include the chromosome operators which shape and guide the algorithm in its execution. Both the fitness function as well as the chromosome operators form key integral components of the genetic algorithm and will be explored in depth later within this paper. 
Current research identified by the authors have all used programs that are "small, in domains with low complexity" which is in contrast with the real world where systems are complex and are made up of various intersecting components. As the problem space of the programs expands and scale upwards with use and the increasing usage and integration of more complex data, further research into GA optimisation has been deemed "necessary" by the authors in order to improve GAs for the generation of test cases of complex data and problems. 

The next review is a SLR into the generation of test oracles through the usage of machine learning by Fontes and Gay \cite{Fontes-MLSys}. The generation of oracles is known as Metamorphic Testing and is a different approach from the automated testing approaches thus far. Nevertheless, metamorphic testing is still a form of test case generation. In the process of creating test cases, a programmer is required to consider standard and edge-case inputs for code. Apart from the actual input specified, the programmer also has to write an "oracle" which "judges the correctness of the resultant execution" which is an effort-intensive task. 
Due to the amount of resources and effort required for manual oracle generation, automated oracle creation is currently of particular interest to researchers. Therefore, a current review of the machine learning approaches to improve and accelerate this process is what the SLR aims to provide. 

Fontes and Gay have attempted to identify the different methods that have been utilised to make the process of generating oracles a lot easier for developers. They have performed comparisons between different papers and have concluded that "machine learning has the potential to solve the test oracle problem" which is the challenge of generating test oracles automatically \cite{Fontes-MLSys}. They have also found that most of the machine learning approaches involved some form of Neural Network as a technique and was used to "train predictive models that served as a stand-in for an existing test oracle". This has allowed for powerful inferences to be generated from program code that would have otherwise not been possible. Fontes and Gay have however also concluded that there are certain elements of weakness that would ideally need to be addressed or require further research towards. 

Fontes and Gay remark that many of the proposed neural network approaches be it "Back Propagation", "Multi-layer Perception", "Probabilistic" and more are "based off on simple neural networks with only a few hidden layers" \cite{Fontes-MLSys}. These techniques are limited in the complexity of the functions that they can model and have already been replaced by newer and better approaches for neural networks. In their identified challenges, the authors also state that the oracle generation is quite limited by the contents of their training data and that after the training process, the models generated by these above neural network techniques will have a fixed error rate. This means that the models will not be able to learn from any new mistakes and improve after the initial training which indicates that it is static and not dynamic in adapting to new circumstances \cite{Fontes-MLSys}. Fontes and Gay therefore conclude that the models should be retrained over time and better training data needs to be sourced to improve the models used for test generation. Finally, in a theme that appears to be reoccurring throughout various papers, the authors also identify a challenge where various papers are making conclusions based on examples that are too simplistic in nature which is not representative of real-world applied usage. They also note the lack of a common benchmark for test case generation to compare approaches to which is a similar concern raised by Anand et al. and Ali et al. alongside other researchers within this field. More research into the usage of machine learning techniques is required and we will also explore this ourselves later as we discuss improvements made to SBST. 

The final literature review is a comparison review by Kurmaku et al. \cite{Kurmaku-HumanTest-Lit} between human-designed and written test cases as compared to automated approaches. In their analysis, the authors report that most of the studies "do not report a significant difference between manual and automated test generation" due to the perceived advantages and disadvantages of both approaches. A reoccurring fact is that automated testing is superior over manual testing due to its ability to produce large amounts of test cases in a short period of time very quickly. Kurmaku et al. \cite{Kurmaku-HumanTest-Lit} also notes that automated testing can achieve a higher coverage score and account for edge cases that a developer may miss during manual testing. Despite this however, they also note that manual testing "may achieve higher mutation and fault detection scores" and that the "vast majority of papers report that automated testing are slightly worse in fault detection than manual testing". This indicates that manually created tests were able to identify issues wrong with a system under test more readily and also be more targeted. They also note that automated testing suffers from poor readability which may be an impediment to the developers attempting to understand the automated code which may add on to their respective workloads. Whilst their literature review appears to suggest that automated testing is superior, the authors have found that in some cases manual testing may still be superior but have called for the need for more analysis to form a more objective view. The authors also conclude that both manual testing and automated approaches will be superior over random testing in terms of the faults detected alongside general code coverage. This call for more analysis is a common request as seen from previous literature reviews above. 

As discussed earlier within our justification, we can therefore see that the previous reviews above have explored GAs in some depth. They have however not explored the various improvements that researchers have made to the underlying GA approach or the various problems that face GA, PSO and AUTG in general. We therefore present this systematic literature review to address those gaps and provide the reader with a contextual overview. 


\section{Methodology} \label{chap:methodology}


\subsection{Filtering Methodology}

As a systematic literature review, this paper based on selection and exclusion criteria, will consolidate papers that are relevant towards our research questions as well as the scope that we have defined. The most important aspects of the filtering methodology are the "transparency" of the process as well as its "reproducibility". Booth et al. \cite{Booth-SysReview} discuss how a systematic review "allows for a trustworthy answer that is based on the whole truth from all the evidence" rather than simply just a "mind-numbing list of citations and findings that would resemble a phone book". This review would also enable a new interpretation in existing research and follows the following reproducible steps as follows:

\subsubsection{Stage 1 Filtering - Initial Selection}\label{Stage_1_Filtering}
\mbox{}\\
In order to obtain all the relevant papers for our research questions and fundamental topic, a series of searches utilising certain keywords related to "test case generation" was entered into four databases which were: 

\begin{itemize}
    \item Scopus
    \item Web of Science
    \item IEEE Xplore
    \item ACM Digital Library 
\end{itemize}

These four databases were selected due to their renown within the academic world. IEEE Xplore contained a multitude of papers pertaining towards computer science and engineering research as well as various conference articles about innovations regarding technology. ACM Digital Library was also chosen due to its main focus in computing literature. All of the databases had an extensive catalogue of peer-reviewed academic papers. These databases afforded us the relevant papers needed for our review. 

The following keywords were utilised within these databases to obtain a series of papers that would be further filtered below. Each keyword was entered into each databases' respective search area to which the resultant listings were captured and downloaded. The following keywords that were utilised are listed in Table~\ref{table: keywords}:

{
    \renewcommand{\arraystretch}{1.5}
    \begin{table}
        \centering
        \caption{List of Keyword Searches Utilised in Database Searches}
        \label{table: keywords}
        \begin{tabular}{cc}
            \toprule
            KEYWORD ID & QUERY \\
            \midrule
            KEYWORD ID 1 & "unit test code generation" \\
            KEYWORD ID 2 & "unit test code generator" \\
            KEYWORD ID 3 & "unit test code automation" \\
            KEYWORD ID 4 & "unit test generation" \\
            KEYWORD ID 5 & "unit test generator" \\
            KEYWORD ID 6 & "unit test automation" \\
            KEYWORD ID 7 & "unit testing generation" \\
            KEYWORD ID 8 & "unit testing generator" \\
            KEYWORD ID 9 & "unit testing automation" \\
            KEYWORD ID 10 & "test case generation" \\
            KEYWORD ID 11 & "test case generator" \\
            KEYWORD ID 12 & "test case automation" \\
            KEYWORD ID 13 & "assertion generation" \\
            KEYWORD ID 14 & "assertion generator" \\
            KEYWORD ID 15 & "assertion automation" \\
            KEYWORD ID 16 & "automatic software testing" \\
            \bottomrule
        \end{tabular}
    \end{table}
}

Note that our keywords were very broad in search and effectively retrieved all papers that were related to unit test case generation. Some papers utilise "assertions" over "unit test" which is similar in meaning and is thus why we also included that search. It is apparent that we could have used a different search approach or the PICO approach that was utilised by Kurmaku et al. within their review which would allowed for more specific targeting and filtering of papers for review. Whilst very broad, the 16 keyword searches allowed us to capture all of the research papers and articles related to Automatic Test Case Generation. This allowed us to be sure that we were not missing out on any paper that may have inadvertently been filtered out as a result of a more exclusive search process. 

The broad parameters of the search and retrieving all possible papers also highlights the "all-encompassing" nature of this Systematic Literature Review. From this initial search, a total of \textbf{10,737} papers were found as a result of the search queries above. After factoring out the duplicate papers that were overlapping as a result from being published into multiple databases, a final count of \textbf{5300} papers that satisfied the first criteria was available at our disposal to use. The table containing the paper count from each respective database can be seen in Table~\ref{table: FinalCount}. The searches and papers were performed and retrieved from the respective databases in February of 2023. 

{
    \renewcommand{\arraystretch}{1.5}
    \begin{table}
        \centering
        \caption{Final Paper Count}
        \label{table: FinalCount}
        \begin{tabular}{cc}
            \toprule
            Database & Paper Count \\
            \midrule
            Scopus & 3844 \\
            Web of Science & 2241 \\ 
            IEEE Xplore & 1585 \\
            ACM Digital Library & 2474 \\
            \midrule
            Total & 10,737 \\
            Final Total (No Duplicates) & \textbf{5300} \\
            \bottomrule
        \end{tabular}
    \end{table}
}

\subsubsection{Stage 2 Filtering - Abstract Filtering}
\mbox{}\\
The next stage of the process was to perform filtering on the final 5300 different papers that were received. Abstract filtering is the process of reading through the title and abstract of a paper in an effort to separate between the papers useful to the topic at hand away from the other papers that may not be as relevant. 

As the goal of this Systematic Literature Review is to compare studies generating test code automatically, selection criteria that matched that goal was applied to filter through the 5300 papers. More precisely, papers were selected based on the following inclusion and exclusion criteria:

\begin{itemize}
    \item Papers must be written in English.
    \item Papers must have been released after 1990 on-wards. 
    \item Papers must have been peer-reviewed (Conference Papers and Journals).
    \item Must have some form of machine learning methods/algorithms to generate the test cases and assertions.
    \item Must be white box testing. 
    \item Must use source code as inputs and not use any form of UML diagrams, Model-based Testing or Natural Language requirements. 
    \item Papers must also generate unit test code as outputs and not any plain text unit test descriptions.  
\end{itemize}

All of the papers that matched the initial keyword searches shown in Table~\ref{table: keywords} had their names, authors and abstracts downloaded into an excel document. Duplicates were removed from the Excel spreadsheet at this stage. The title and abstract of each paper was then read to determine its relevancy. Papers were selected on the basis of whether or not they were similar to the exclusion criteria above and if they were relevant to the research questions identified earlier above. If a paper was found to be suitable, it would be marked as "accepted" in the excel document whilst irrelevant papers were marked as "rejected".

In order to reduce selection bias, another academic researcher (Ario Birmiawan Widyoutomo) also performed the same filtering methodology to mark papers as accepted or rejected. At the end of the abstract filtering step, our differences were compared, and it was determined that for the purposes of continuity and ensuring that nothing was missed, papers that had a conflict would be accepted. This therefore meant that a paper that had a marked conflict on whether to accept or reject between the two researchers would ultimately be accepted. Having another person also perform the filtering step reduces selection bias and reduces the chance for a potentially relevant and useful paper to "slip through the gap" and thus ensures that the review is as systematic and "all-encompassing" as possible. 

At the conclusion of this filtering step, a final total of \textbf{167} unique papers out of the original \textbf{5300}. These \textbf{167} papers represented all of the papers that were relevant to test case generation and that matched our final exclusion criteria. During the filtering step, a lot of knowledge about the field of test case generation was also attained simply through reading the various abstracts. 

\subsubsection{Stage 3 Filtering - Consumption and Analysis}
\mbox{}\\
The next step of the filtering process was to now read through the final \textbf{167} shortlisted papers in order to increase our own knowledge of the topic at hand. This filtering step also allowed us to determine the individual relevancy of the paper to our research goals. Due to our filtering step earlier, we envisioned that most of the papers that we reviewed at this step would also be accepted for use in our systematic review. However, this turned out not to be the case with \textbf{104} papers not being relevant. At the time of the second filtering step, the title and abstract indicated a potential fit. However, the paper would generally end up discussing another concept that was either not relevant to test case generation or for another approach that was out of scope for our review and our research goals. 

A final amount of \textbf{63} papers was ultimately accepted for our systematic literature review and were further tagged and sorted based on their topic at hand. Papers discussing GA and PSO were tagged and grouped together for easy access whilst other relevant topics to our review such as challenges and limitations were also grouped similarly. Whilst we believe that we have filtered and selected a suitable range of papers to use in our review, we believe that it is also important to note that the stage 2 and especially the stage 3 filtering steps was entirely a subjective process. The acceptance and rejection of various papers were exercised at our discretion and was based on our interpretation of the exclusion criteria. This means that another researcher who employs the same filtering methodology outlined above may in fact have a different interpretation on what the final papers to be used in the review should be. Nevertheless, we believe that we have made appropriate decisions in our filtering methodology. 

\begin{figure}
    \centering
    \includegraphics[width=1\linewidth]{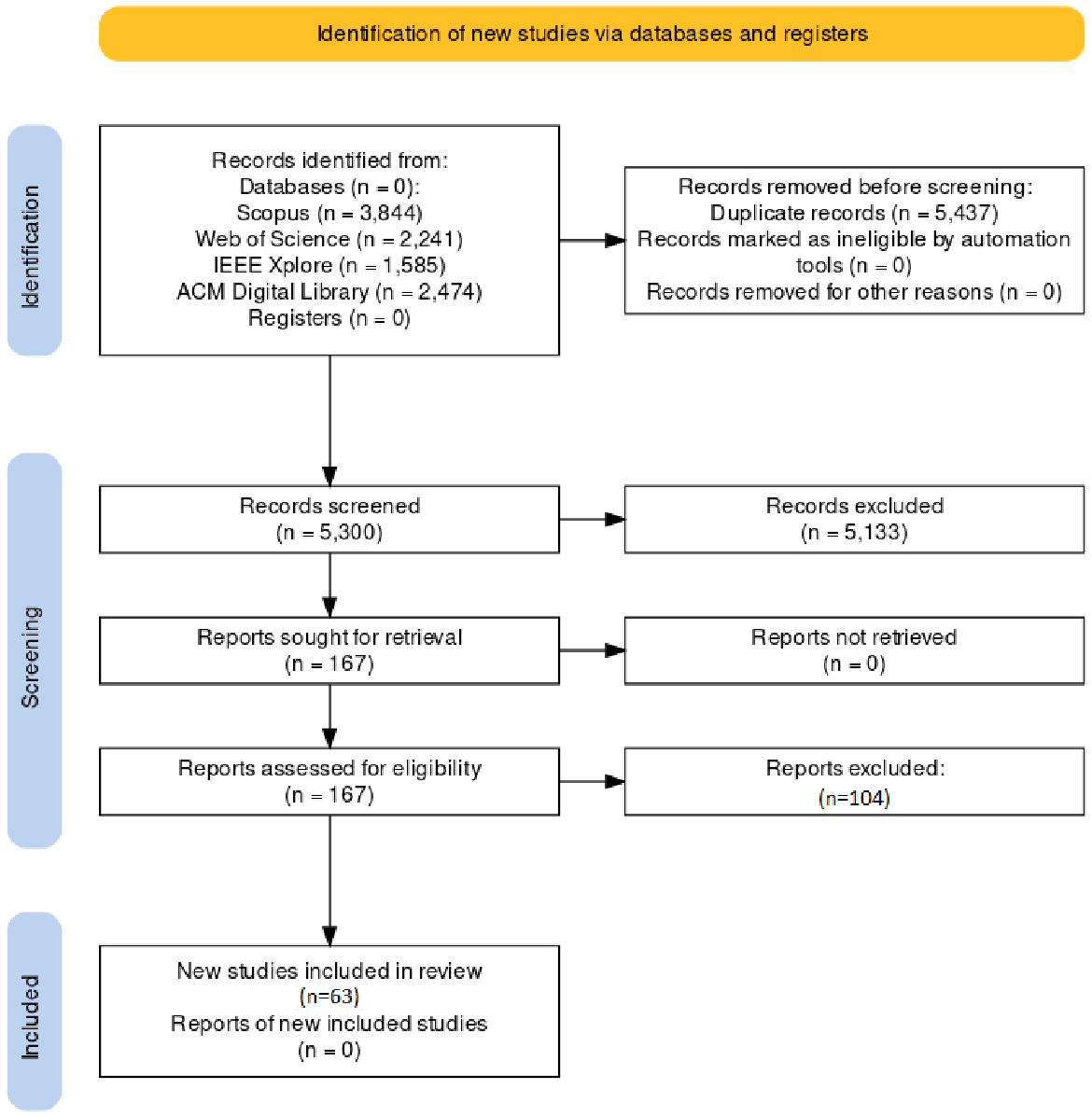}
    \caption{PRISMA Diagram of Selection and Filtering Process}
    \label{fig:Prisma}
\end{figure}

These final papers will be examined further and the next few sections will discuss the findings made by these accepted papers. We have also provided a PRISMA chart outlining our discovery and filtering process as seen in Figure~\ref{fig:Prisma}.

\section{Testing Criterion} \label{chap:adequate testing}

Before we delve deeper into the inner workings of the various techniques of automated test case generation, we need to better understand the concept of "test criterion". The ideal outcome of software testing is to ensure that the software under test runs as intended without any faults. If the size of the program was small enough, it could potentially be possible to test and account for all possible events and ensure that the test suite is robust. This approach however, quickly becomes infeasible and computationally outright impossible with large complex programs that are made up of multiple different components. These programs could have potentially infinitely amounts of different events that cannot be exhaustively tested against. Developers also do not possess the time and resources to continuously test different suites and inputs to ensure the code's behaviour. This thus calls for a comparable and quantifiable metric that can be used to determine the "effectiveness" of a test suite. 

Coverage criterion is therefore used to address this issue and is commonly used within white-box testing. Sun et al. define test adequacy criterion as a method to "quantify the degree of adequacy to which the software had been tested by a test suite using a set of coverage conditions" \cite{Sun-StructuralDNN}. Whilst Sun et al. also argue about the effectiveness of the extent the coverage criteria is able to ensure correct functionality, they state that a test suite that exhibits high coverage would be able to increase the overall confidence in a program. \cite{Sun-StructuralDNN}. The "criterion" thus allows developers to obtain a usable and comparable metric to see the effectiveness of their test suite. It also allows the developer to determine how much of their software has been tested and accounted for. 

\subsection{Criterion}

There are a plethora of different "criterion" that can be deployed to test code. Each have their respective pros and cons and are applicable to different scenarios. Below we have outlined the most common criterion used within automated testing namely the "coverage" family, mutation fault-based testing as well as MC/DC as the focus is to ensure the "correctness" of the system under test. There are also different criterion that are focused on system specifications, performance testing such as threading and race-conditions and many more that have not been mentioned. 

\subsubsection{Statement Coverage}
\mbox{}\\
The first coverage-based criterion we will explore is also the simplest criterion. Statement Coverage is the coverage criterion where a percentage score is given based off the amount of lines (statements) executed. Statement Coverage has been described by Zhang et al. as the "requirement for all of the statements in the program code under test to be executed at least once by the test suite" \cite{Zhang-Smartunit}. The criterion has been identified to be the easiest and most common test criterion. A 100\% code coverage is generally easy to achieve and any score other than 100\% would indicate to the developers that there are certain aspects of code that have not been tested against a particular test case. However, Statement Coverage as a test criterion should not be used on its own as the criterion doesn't necessarily detect program faults - only if it has been executed \cite{Zhang-Smartunit}. 

\subsubsection{Branch Coverage}
\begin{figure}
    \centering
    \includegraphics[width=0.6\linewidth]{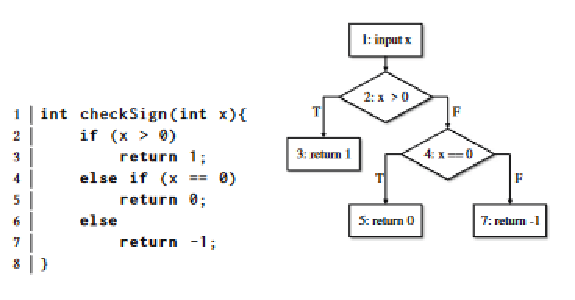}
    \caption{checkSign function showcasing an example decision tree. \cite{Zhang-Smartunit}}
    \label{fig:checksign}
\end{figure}

\mbox{}\\
Branch Coverage ensures that every decision point within a program has been executed and tested for both "true" and "false" values. This means that each of the decision points within a program normally at the if/else statements should be tested for both conditions. In the example as seen in Figure~\ref{fig:checksign}, each branch condition found within line 2, 4 and 6 would need to be checked to see how the program would operate with "true" and "false" branch values. 

 Zhang et al. reiterates this by confirming that the Branch Coverage criterion will "confirm that all of the possible branches from each decision are executed at least once" \cite{Zhang-Smartunit}. The researchers also believe that Branch Coverage is a stronger coverage criterion over Statement Coverage and like the later, is also easy to achieve 100\% coverage within. Whilst overall a simple coverage criterion, Guo et al. have raised concerns however on the rising complexity and increase in time required to perform large-scale branch coverage testing on software systems that are increasingly complex and built up of multiple different components \cite{Guo-AUTG-GAN}. As the complexity of software continues to expand, this issue will continue to propagate and will bring developers back to the original fundamental issue of time and resource-consuming software testing. 

\subsubsection{Path Coverage}
\mbox{}\\
The final testing criterion part of the "coverage" family is Path Coverage. Path Coverage is a criterion that is more strict than Branch Coverage and requires that every possible potential route or "path" within a program to be accounted for. First introduced and used by Howden in 1976, Path Coverage testing is the strongest coverage criteria amongst all coverage-based testing methods as it can detect about 65\% of faults and defects presented within a system under test \cite{Bala-Evolutionary-Path-Coverage}. Mishra et al. continues to advocate for Path Coverage's effectiveness through its ability to also attain 100\% branch and statement coverage as a result of testing every possible control path within the program \cite{Bala-Evolutionary-Path-Coverage}. This is possible in Path Coverage and not in the other coverage methods due to the way all of the unique paths are acquired from a Control Flow Graph (CFG) that outlines all of the unique paths or "flows" that a program can take. 

Despite being a powerful test criterion, Path Coverage testing faces a significant issue where it cannot detect all of the potential errors within a system under test. Mishra et al. earlier gave an average figure of around 65\% detection which is far from the 100\% detection expected. Ji et al. however concludes that it is most likely infeasible and unrealistic to expect 100\% detection. They determine that the "all-paths" nature of Path Coverage testing is "not suitable for complex programs which may have an infinite number of paths" \cite{Ji-DataFlowTesting}. A similar issue to Branch Coverage above and to testing in general, the effort required to account for all possibilities grows exponentially the more complex the software becomes. It is not an understatement to say that there could potentially be infinite paths within the system which makes it impossible to test for all of the unique paths. 

Another group of researchers, Zhang et al., share a similar sentiment and state that it is impossible to achieve total path coverage for complex programs \cite{Zhang-K-Means-Clustering}. They also observe that the selected paths within the program chosen as the target paths will have a direct effect on the quality of the test suite. This means that if the paths within a program were only related to the functionality of a particular component then that component would be readily tested at the detriment of the other components within the program. Zhang et al. proposed a solution to this issue which would still attain high path coverage through the usage of "K-means Clustering". The algorithm in question would group paths by similar characteristics into "clusters". The paths at the centre of those clusters would then subsequently be chosen as "target" paths for testing and allow "different areas of the program to be tested" \cite{Zhang-K-Means-Clustering}. Overall, whilst not achieving a full coverage, the usage of this algorithm has increased path coverage and the researchers have found it to be better as compared to a normal random selection.

\subsubsection{Mutation Score}
\mbox{}\\
Whilst not part of the "coverage" family, another criterion in use by developers and embedded within testing methods is Mutation Testing. Mishra et al. defines Mutation Testing as a "fault-based white-box software testing method" where "artificially generated faults are seeded into software through some predefined mutation operators" \cite{Mishra-Genetic-MT}. The test criterion evaluates the effectiveness and quality of the test data by generating active "mutants" within the code and viewing whether or not the test suite is able to identify those mutants. As articulated by Jatana et al., "the underlying principle of mutation testing is to emulate faults in software that a proficient programmer may make during the software development phase" \cite{Jatana-PSO-MT}. A comparable metric known as the Mutation Score is determined at the end based on the amount of mutants that were detected and eliminated by the test code. Mutation Testing is usually performed after coverage-based testing is completed to ensure that the system under test is covered 100\% by the test code and so that certain functions aren't missed \cite{Mishra-Genetic-MT}.

\begin{figure}
    \centering
    \includegraphics[width=1\linewidth]{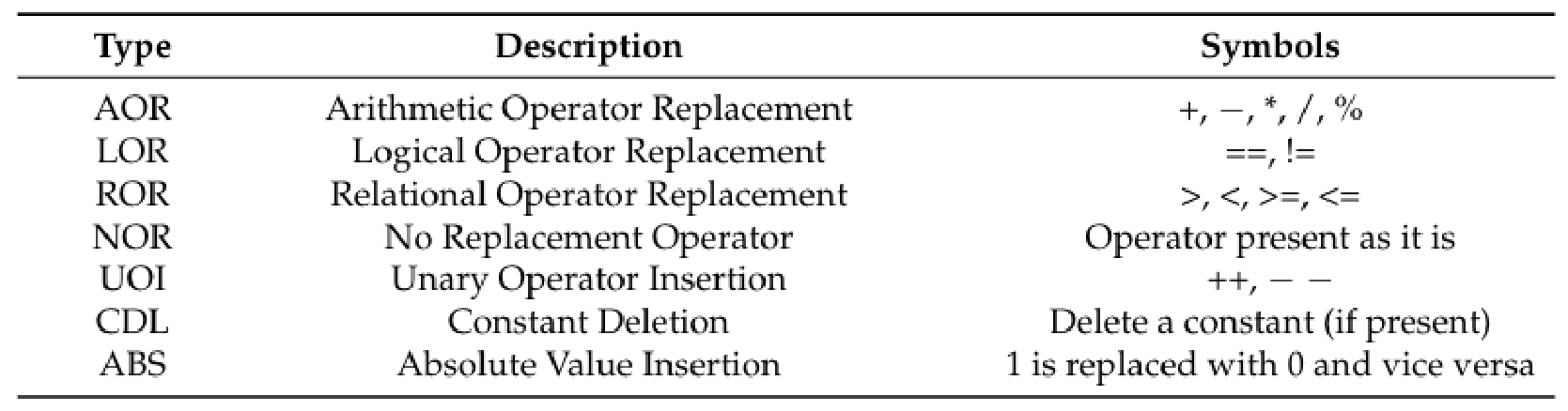}
    \caption{Different operators used for mutation analysis \cite{Mishra-Genetic-MT}}
    \label{fig:mutation_operators}
\end{figure}

\begin{figure}
    \centering
    \includegraphics[width=0.85\linewidth]{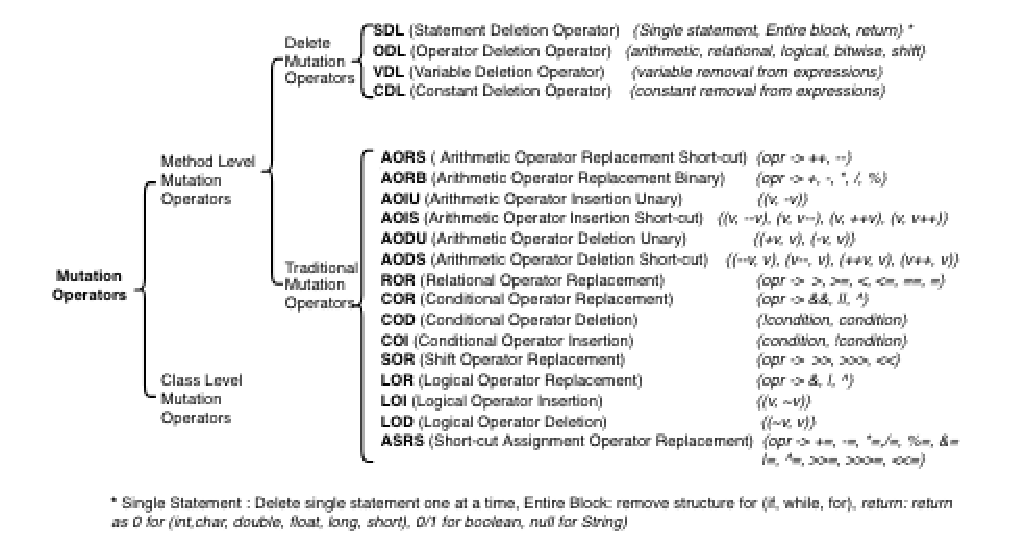}
    \caption{More Mutation Operators \cite{Rani-Elitist-GA}}
    \label{fig:more_mutation_operators}
\end{figure}

At the conclusion of the coverage test, the system under test is tested with the test cases that were generated as a result of the coverage test. This step leads into mutation testing where "certain mutation operators are inserted into the code to "mutate" the program". These mutation operators vary in size and function but all aim to change the program from its original state. An example of the set of mutation operators used can be seen in Figure~\ref{fig:mutation_operators}. A more extensive list of mutation operators can likewise be viewed in Figure~\ref{fig:more_mutation_operators}.

The mutated program is then subsequently tested using the original generated test cases. A test case that is deemed good and valuable to the developer is a test case that can accurately recognise that the program has been changed and thus has induced errors. If a test case is able to detect a mutant, the mutant is declared "killed" and is accounted for. Mishra et al. states that the overall goal is to select the most efficient test cases that are able to detect a large variety of mutants within a system under test. If a mutant is not killed however, it indicates two possible scenarios as envisioned by Rani et al. The first scenario is that the generated mutants are not different from the original program under test and therefore the mutated program is actually identical (or slightly different). This could potentially be as a result of the mutation set that was used to generate the mutants. The other scenario is that the generated test suites were inferior in nature and were not able to detect the mutated faults present within the program \cite{Rani-Elitist-GA}.

\begin{figure}
    \centering
    \includegraphics[width=0.5\linewidth]{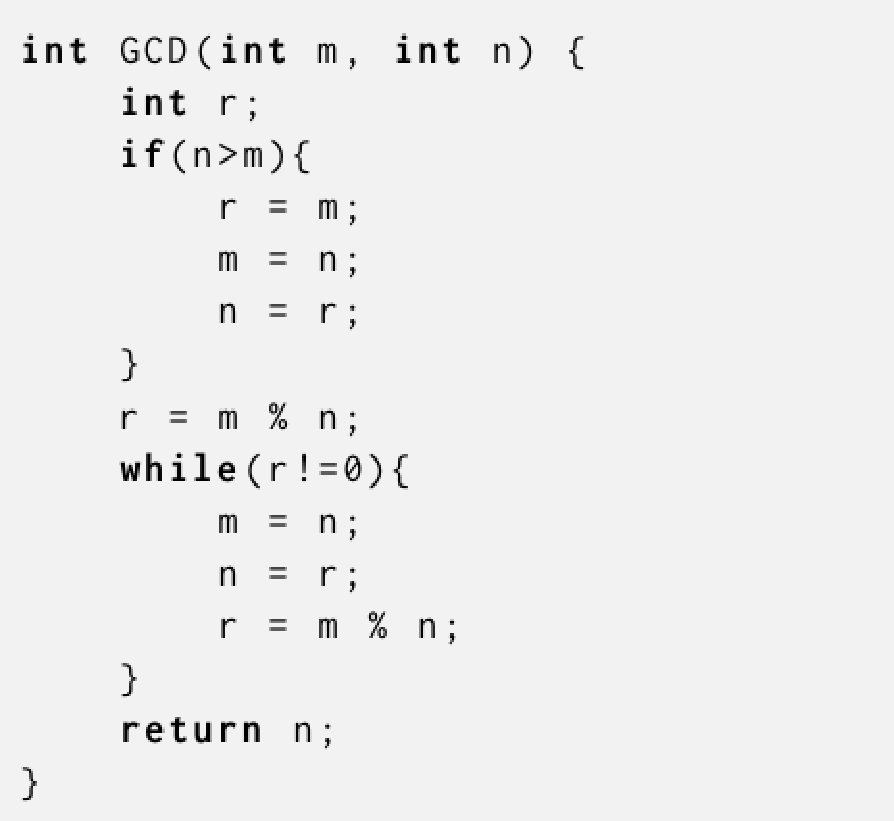}
    \caption{Source code of GCD program \cite{Mishra-Genetic-MT}}
    \label{fig:GCD}
\end{figure}

\begin{figure}
    \centering
    \includegraphics[width=1\linewidth]{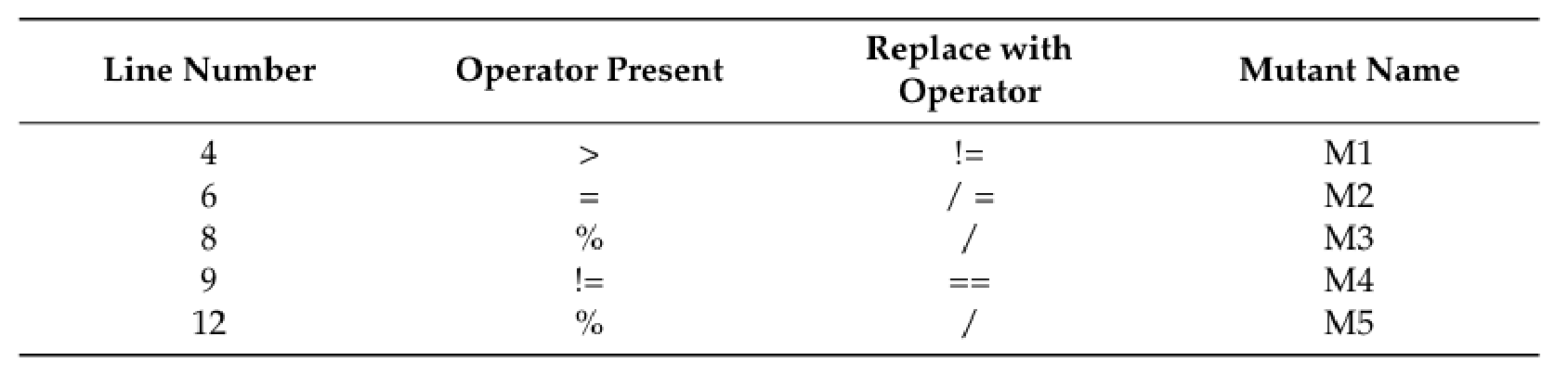}
    \caption{Operators used in the GCD for mutation testing \cite{Mishra-Genetic-MT}}
    \label{fig:GCD_Operators}
\end{figure}

The mutation process can be seen in the example code listed in Figure~\ref{fig:GCD}. The program is known as the Greatest Common Divisor (GCD) and produces the greatest common divider of any two input integers. As seen in the Figure~\ref{fig:GCD_Operators}, the mutation operators will purposely change the original operators in an effort to see whether or not the original test cases can "detect" the mutated elements and thus "kill" the mutation. The number of mutants a particular test case is able to detect and kill indicates the effectiveness of the test case. 

It is important to note here that we have a preference for a smaller optimised set of test cases that can accurately test for an entire code-base rather than a plethora of small test cases. The more test cases there are, additional complexity is added and it becomes more likely for test cases to "double-up" in functionality and become redundant and thus add to the developer's workload as they attempt to understand the repetitive test cases. The efficiency of the mutation testing process is therefore assessed through the following formula as seen below in Figure~\ref{fig:mutation_score}. Within this figure, "MS" refers to the mutation score, the "Total Mutants Killed" refers to the amount of mutants detected by the original test cases and finally the "Total Mutants Present In The Program" refers to the amount of mutants present in the program created by the mutation operators earlier. The above represents a simplistic overview of how Mutation Testing is performed to which the Mutation Score is then utilised to compare different test suites for effectiveness. 

\begin{figure}
    \centering
    \includegraphics[width=0.5\linewidth]{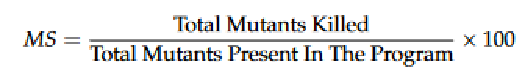}
    \caption{Mutation Score Formula \cite{Mishra-Genetic-MT}}
    \label{fig:mutation_score}
\end{figure}

There has also been a great deal of work by multiple researchers looking into optimising the mutation process to achieve a higher score. Estero-Botaro et al. reflects that it is relatively easy to obtain mutation scores within the 50\%-70\% range but also acknowledges that getting that score to 100\% is currently a very hard task and is presently an active field of study for many researchers within the field \cite{Estero-WSBPEL}. So how does one increase mutation score? The first idea is to simply remove the "poor-performing" test cases that are only able to detect and eliminate one mutant within the program and to keep the test cases that could detect and eliminate multiple mutants. This is however, a bad approach and Estero-Botaro et al. emphasises the "number of mutants killed per test case" should not be used as an appropriate fitness metric \cite{Estero-WSBPEL}. The reason for this is that there could exist specialised test cases that are niche and specific enough to identify and eliminate a particular mutant. The removal of that test case would have resulted in a much lower average mutation score. 

In regards to optimisations, Du et al. have proposed a subset of mutators that can reduce the total amount of mutants from 19 operators and down to just 5 whilst still retaining a 95\% variance score \cite{Du-MT-Optimisation}. The reduction in the amount of operators will reduce run time complexity and reduce potential double-ups with other operators. The five operators are as follows: Absolute value Insertion (ABS); Arithmetic Operator Replacement (AOR); Logical Connector Replacement (LCR); Relational Operator Replacement (ROR); and Unary Operator Insertion (UOI). These mutation operators execute actions such as changing the arithmetic expressions to positive values, flipping relational signs such as ">=" or "!=" as well as changing conditional checks. Mishra et al. have also put forward their improvement for mutation testing which involves combining path coverage with a fault detection matrix (FDM). The path coverage test data is initially stored in an optimised test suite to which it is then executed and used to kill all of the generated mutants within the system under test. The goal of this optimisation is to minimise the amount of test cases that essentially perform the same role and provide the same amount of information as another test case. The reductions in test cases however still maintains the same mutation score for the entire test suite throughout the process and does not conflict with the findings stated earlier by Estero-Botaro et al. Overall, 
Mutation Testing is a powerful technique and has been integrated within other automated methods such as the Genetic Algorithm. Later within this paper, we will explore how other algorithms have used Mutation Testing and its criterion. 

\subsubsection{MC/DC}
\mbox{}\\
The final criterion that we would like to explore is known as Modified Condition / Decision Coverage (MC/DC). MC/DC is a criterion that was developed by NASA and is used in the aviation industry to meet DO-178C aviation standards \cite{Sun-StructuralDNN}. The criterion is used for the testing of applications that are critical in its operations and can not fail such as aircraft. Sun et al. states that the primary concept behind MC/DC is that "if a choice can be made, all of the possible factors (conditions), that contribute towards that choice (decisions) must be tested" \cite{Sun-StructuralDNN}. This is to ensure that every possible condition and value has been accounted for and tested prior to its deployment. The following decision expression and its associated test cases represent an example of MC/DC \cite{Sun-StructuralDNN}:

$((a > 3) \vee (b = 0)) \wedge (c \neq 4)$

$(1) (a > 3)=false, (b = 0)=true, (c \neq 4)=false$

$(2) (a > 3)=true, (b = 0)=false, (c \neq 4)=true$

$(3) (a > 3)=false, (b = 0)=false, (c \neq 4)=true$

$(4) (a > 3)=false, (b = 0)=true, (c \neq 4)=true$

Sun et al. raise the point that although the first two cases already account for both decision and condition converage, the final two cases are required as MC/DC stipulates that every condition should evaluate to true or false at least once. 

As compared to the other criterion, Song et al. believes that MC/DC coverage strength is a lot higher than the other coverage criterion due to its stringent requirements. Song et al. however also state that the criterion can be disadvantageous similar to Branch Coverage where "when there are too many conditions in the branch node, it [the process] will excessively consume much time" \cite{Song-MC/DC}. These findings are also reaffirmed by Zhang et al. to which they state MC/DC has a stronger coverage criterion over Branch coverage but that in practice, "MC/DC is usually difficult to write manually because of the complex logic in decision condition" \cite{Zhang-Smartunit}. 

\subsection{Feasibility}

The above criterion all represent criterion that have been utilised white-box software testing. They have been used as part of methods to generate test cases automatically and have been used to compare test suites. The above only represents a small subset of criterion that are out there. In terms of feasibility, Path Coverage and Mutation Score are commonly used but all suffer from the potential issues that comes with a large program due to the potentially infinite amount of "paths" that could exist including cyclic loops. Path Testing however represents the most "robust" criterion that is generally used for coverage and fault detection in addition to Mutation Testing. As a result of this section, the foundations  of automated testing are established and the reader should have now attained an understanding of the "criterion" that are in use by automated methods. 


\section{Automated Approaches} \label{chap:automated approaches}


\subsection{Random Testing}

Prior to our dive into the evolutionary methods, we would like to present some context to the reader by examining the out of scope black-box approach known as Random Testing. Anand et al. defines Random Testing as "one of the most fundamental and most popular testing methods" as it is simple in context and easy to execute \cite{Anand-Orchestrated-Survey}. As Random Testing is a black-box method, it is unable nor does it care about the internal functions that a software under test possesses and is thus only affiliated with the inputs into a program. As a result of this, Anand et al. conclude that Random Testing may be the only feasible method if the specifications of a given program to be tested are not present or if the source code is unavailable \cite{Anand-Orchestrated-Survey}. Liu and Yu explain that inputs are selected at random and tested to see if the software executes as intended. The process will also randomly build upon large amounts of test sequences and attempts to test various combinations of function calls randomly \cite{Liu-Randoop-TSR}. Random Testing has its advantages where it is relatively easy to setup and execute without needing anything extra. It however has disadvantages in which it may generate a large test suite of redundant and hard to read test cases whilst also taking a long period of time to generate. These disadvantages are explored below in a comparison between search-based techniques. Random Testing has also been applied into an AUTG tool that is currently used in practice known as Randoop which is "one of the most used automated tools based on Random Testing in academia" \cite{Liu-Randoop-TSR}. It has also been improved upon through the implementation of Adaptive Random Testing (ART) and MoesART which are explored below. 

\begin{figure}
    \centering
    \includegraphics[width=0.8\linewidth]{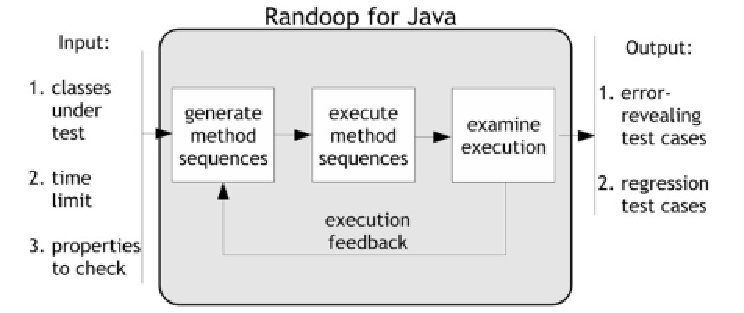}
    \caption{Randoop Overview \cite{Liu-Randoop-TSR}}
    \label{fig:Randoop}
\end{figure}

\subsection{Search-Based Evolutionary Testing}

In the following section we present two search-based testing methods of interest which are the Genetic Algorithm as well as the Particle Swarm Optimisation Algorithm. Sharma and Pathik state that search-based testing utilises heuristic information for the automated generation of test cases. These test-cases are then minimised to ensure that only the ones required are kept in order to satisfy the objective fitness function \cite{Sharma-CSA}. Whilst we will only be focusing on two meta-heuristic search algorithms today, there have been a wide range of other algorithms presented in research such as: Ant Colony Optimisation; Monarch Butterfly Optimisation; Moth Search Algorithm; Crow Search Algorithm; Harris Hawks Optimisation; RUNge-Kutta method; and much more \cite{Mishra-Genetic-MT}.

\subsubsection{Evolutionary Genetic Algorithm}
\begin{figure}
    \centering
    \includegraphics[width=0.70\linewidth]{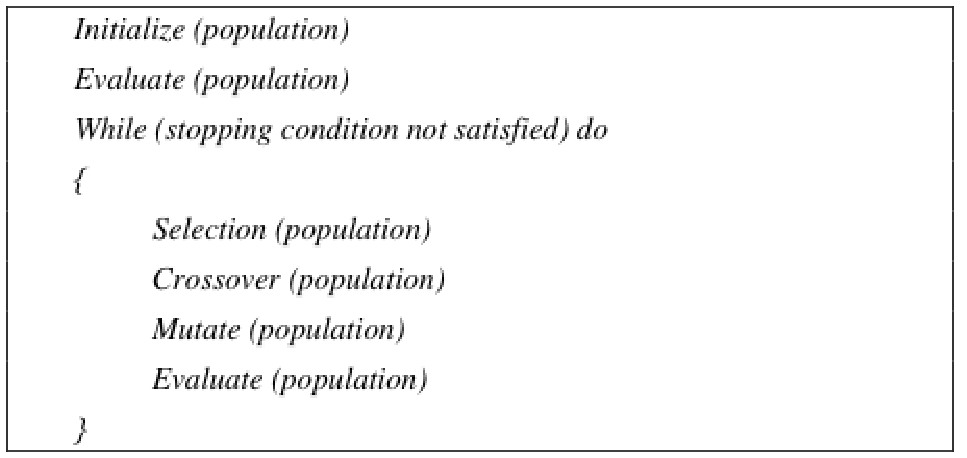}
    \caption{Genetic Algorithm - Pseudo Code \cite{Rao-GAOptimise}}
    \label{fig:GA-Pseudo}
\end{figure}

\mbox{}\\
Genetic Algorithm has been defined as a "probabilistic search technique based off the theory of evolution and natural selection by Charles Darwin" \cite{Estero-WSBPEL}. First proposed by Holland in the early 1970s, its primary purpose is to develop a solution based on the concept of "survival of the fittest" proposed by Herbert Spencer \cite{Rao-GAOptimise} \cite{Estero-WSBPEL} \cite{Khan-GAPathTesting}. Estero-Botaro et al. defined GAs as a "selection process" where the best individuals from each successive generation of a population are kept, mutated and recombined to form a new superior population. The authors also state that there is no "single type of GA" as there are multiple different implementations that are dependent on the objective with the fitness "measuring the quality of the solution" \cite{Estero-WSBPEL}. Despite the various differences a particular GA approach may have from another GA approach, there is still a common flow that all GA methods follow as shown by the pseudo code example in Figure~\ref{fig:GA-Pseudo} by Rao et al. 

Rodrigues et al. explains the process that the GA will take in the following steps \cite{Rodrigues-SysMapping}: 
\begin{enumerate}
    \item The initial population of "n" chromosomes which represents the individuals of the population are first randomly generated. 
    \item Depending on the objective, a fitness function will assign a fitness value to each individual chromosome within the population. 
    \item The following steps are repeated until "m" new chromosomes are generated"
    \begin{enumerate}
        \item The selection operator will choose two chromosomes to reproduce.
        \item The crossover operator will combine elements of both chromosomes together with a randomised probability. 
        \item The mutation operator (if needed) will also further alter the offspring of the chromosomes. 
    \end{enumerate}
    \item The replacement procedure will determine which individuals will be kept for the next generation - thus designating them as the "fittest" and "best" individuals to be kept. 
    \item If the new population still does not match the conditions of termination, continue with the process from the second step. 
\end{enumerate}

\begin{figure}
    \centering
    \includegraphics[width=1\linewidth]{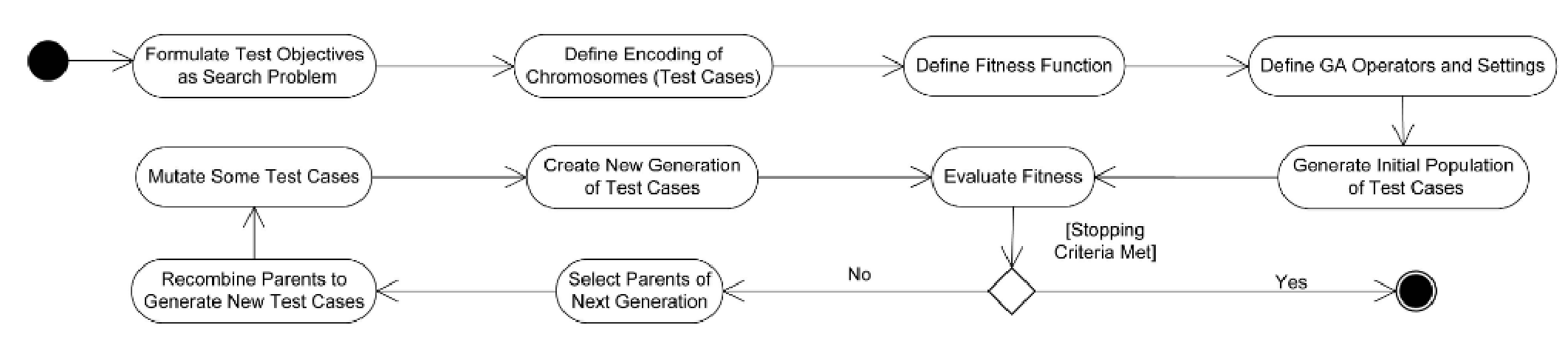}
    \caption{Test case generation flow using Genetic Algorithms \cite{Ali-SysReview}}
    \label{fig:GA-Flow}
\end{figure}

A further flow diagram of the GA test case generation process presented by Ali et al. can be seen in Figure~\ref{fig:GA-Flow}. 

Depending on the objective of the problem, the fitness function will usually attempt to resolve a goal. Otherwise for test case generation, Zhu et al. state that the fitness function will usually be "designed based on path and branch coverage" which are generally the most extensive coverage criterion as we explored earlier \cite{Zhu-Improved-GA-Path}. Within the process itself, Estero-Botaro et al. states that the GA will predominately use the selection and reproduction operators. The authors explain that the selection operators will select individuals within a population based on the fitness of the individual with "the higher the fitness of an individual, the higher the probability of being selected" \cite{Estero-WSBPEL}. Within the reproduction operators, Nayak and Mohapatra state that the main purpose of the crossover operator is "to exchange information between the two parent chromosomes to produce offspring for the next generation" \cite{Nayak-PSO-DF}. They also claim that the purpose of mutation is to "introduce genetic diversity into the population". This step and mutation in general is important as the operators will "manipulate the population" and guide the search away from "unpromising areas of the search space and towards promising ones" as stated by Michael et al. Without this step, the GA process may become "stuck" in a local space which will explore further below \cite{Michael-AUTG-Evolution}. An example of the process described by Rani et al. can be seen below in Figure~\ref{fig:GA_Basic_Process}. Within this figure, we can see the initial population representing the integers "42" and "38" undergoing binary crossover and mutation to output a new population of the integers "46" and "44" \cite{Rani-Elitist-GA}. 

\begin{figure}
    \centering
    \includegraphics[width=0.8\linewidth]{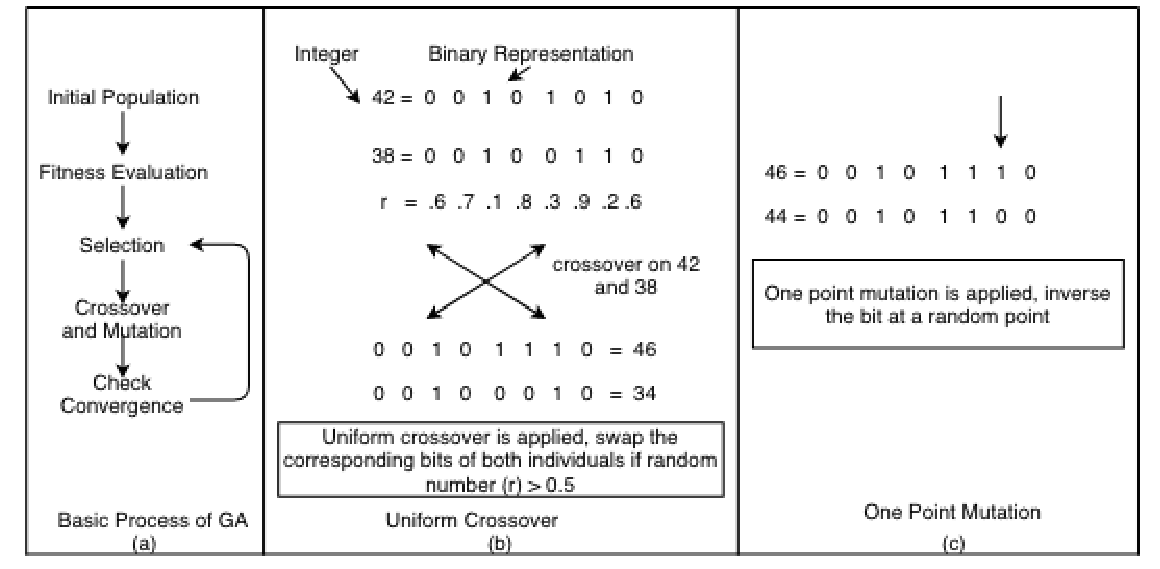}
    \caption{Basics of GA (a) basic process, (b) Uniform Crossover and (c) Mutation \cite{Rani-Elitist-GA}}
    \label{fig:GA_Basic_Process}
\end{figure}

\begin{figure}
    \centering
    \includegraphics[width=0.5\linewidth]{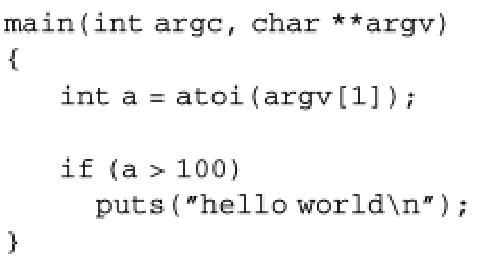}
    \caption{A simple C program \cite{Michael-AUTG-Evolution}}
    \label{fig:GA_C_Program}
\end{figure}

We will now explore a sample execution of the GA AUTG method on a simple C program that was presented by Michael et al. As seen in Figure~\ref{fig:GA_C_Program}, the program will output "hello world" if the value of "a" received from the first command-line argument is greater than the integer "100". 

Within the program, Michael et al. states that the objective function will be based on the value of \textit{a} in line 3 as its value will determine whether or not the if conditional statement is reached or not \cite{Michael-AUTG-Evolution}. The objective is to thus generate populations that satisfy this objective "fitness". When the GA method is utilised, it will generate an initial population randomly as explained by Rodrigues et al. earlier. Michael et al. have assumed an initial population of the values \textit{94, 91, 49} and \textit{-122}. Immediately based off of these values and the resultant condition in line 4, none of the individuals within the population would be accepted. The values generated will also be given a "fitness" value of how close it is to the objective of "100". In this case, the values are given fitness of \textit{6, 9, 51} and \textit{222} which represents numerically the difference that exists between them and the objective. These are subsequently converted into probabilities \textit{1/6, 1/9, 1/51} and \textit{1/222}. The probabilities here will determine the chance of the individual being selected within the population for the next generation. Whilst not really relevant here in a small population of 4, these probabilities have extensive weight in a large population of individuals. As we can see here, the value of \textit{94} has a chance probability of \textit{1/6} which is significantly higher than the probability of \textit{1/222} represented for the value of \textit{-122}. This is that resulting process of GA guiding the search away from the areas of population that will not be relevant as described earlier \cite{Michael-AUTG-Evolution}. 

During the selection process, the highest probability values of "94" and "91" are chosen. As similarly seen in Figure~\ref{fig:GA_Basic_Process}, the values of the two parents will be converted into binary form which are \textit{01011011} and \textit{01100000} respectively as stated by Michael et al \cite{Michael-AUTG-Evolution}. Within the selection process, a random crossover point is chosen which in this example was 3. This meant that the first 3 bits of the binary number was exchanged from the two parents resulting in the offspring \textit{01000000} and \textit{01111011} which are the binary representations of the integers "64" and "123" \cite{Michael-AUTG-Evolution}. This selection and crossover process continues until we have the same amount of off-springs as we did as the starting population. 

In this instance, we now have a valid individual integer \textit{123} which satisfied the objective and thus ends the GA process. The GA process will however continue if the requirements of an objective isn't complete or if the search budget and time for execution is exhausted. As explained earlier, the objective is usually 100\% path coverage but this is generally infeasible to achieve for complex programs. Michael et al. also claims that in the event of a population that even after the crossover process yielded no suitable candidate, the Genetic Algorithm would need to wait for a mutation operator to mutate the population for a different output \cite{Michael-AUTG-Evolution}. Rodrigues et al. states that the mutation operator is responsible for providing genetic diversity within a population and is one of the mechanisms preventing GA from being stuck in a "local optima" \cite{Rodrigues-SysMapping}. The diversity of a population with many different individuals is thus a very important topic which we will continue to explore below. 

In conclusion, this forms the overview of the Genetic Algorithm AUTG method that we had set out to explore in the beginning. It partially answers our first research question that we had declared earlier in the Introduction of this paper. The Genetic Algorithm today is still extensively used in AUTG research and has been used as a base in AUTG tools such as Evosuite alongside others. We will now explore the other popular and effective evolutionary AUTG method known as Particle Swarm Optimisation and will also continue to explore performance improvements of the GA and PSO in later sections. 

\subsubsection{Particle Swarm Optimisation Algorithm }
\mbox{}\\
The evolutionary AUTG method known as Particle Swarm Optimisation (PSO) was first introduced by Kennedy and Eberhart in 1995, roughly 20 years after the Genetic Algorithm was first proposed \cite{Jia-PSO-Test-Generation} \cite{Windisch-Apply-PSO}. The method itself is also an evolutionary process that mimics the "swarming" behaviour present in animals such as birds or fish when they found roosting areas or food sources \cite{Ji-DataFlowTesting} \cite{Bala-Evolutionary-Path-Coverage} \cite{Windisch-Apply-PSO}. Windsch et al. states that the PSO is similar to GA with a population initialised of random "particles". Each particle would "maintain its own current position, its present velocity as well as its personal best explored location so far" \cite{Windisch-Apply-PSO}. The similarities between PSO and GA have been outlined by Jatana et al. to which they state the following \cite{Jatana-PSO-MT}:
\begin{itemize}
    \item A Particle in PSO is analogous to a Chromosome in GA.
    \item The acceleration velocity in PSO is similar in behaviour to the Crossover operation in GA. 
    \item Both PSO and GA conclude upon the termination criteria being met.
\end{itemize}

Windisch et al. explains that the the particles will explore the search space of a problem and will be able to share their "personal experience" to the rest of the swarm. This means that all of the particles within the population will know their personal best locations which is denoted as \textit{pbest} as well as the global best location which is in turn denoted as \textit{gbest} \cite{Windisch-Apply-PSO}. Jatana et al. explains that the PSO will try to solve an "optimisation problem" where the particles are moved around in the search space "by using mathematical formulae based on the particle's current position and velocity" \cite{Jatana-PSO-MT}. The workflow of the PSO method can be seen in Figure~\ref{fig:PSO-Workflow} below. The particles will "learn from different neighbours" and will update their positions and velocities until no more improvements can be achieved. The steps of the PSO method has also been presented by Jia et al. as seen below in Figure~\ref{fig:PSO}.

\begin{figure}
    \centering
    \includegraphics[width=0.4\linewidth]{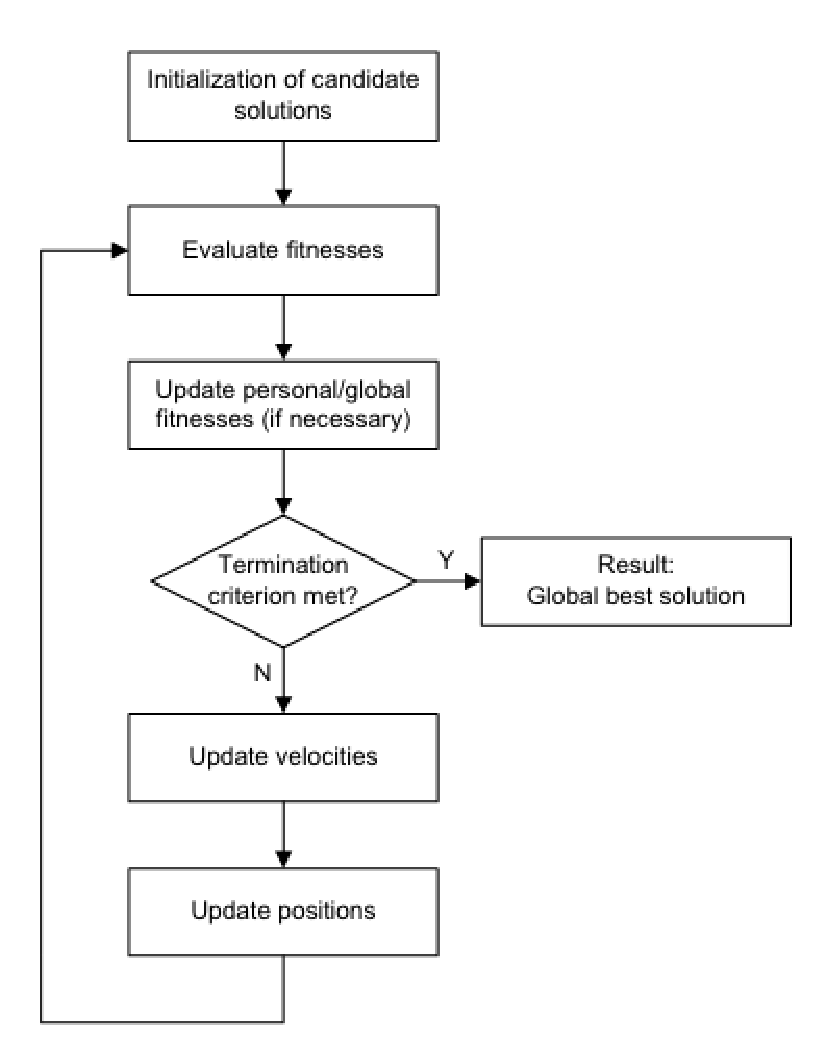}
    \caption{Workflow of a PSO algorithm \cite{Windisch-Apply-PSO}}
    \label{fig:PSO-Workflow}
\end{figure}

\begin{figure}
    \centering
    \includegraphics[width=0.85\linewidth]{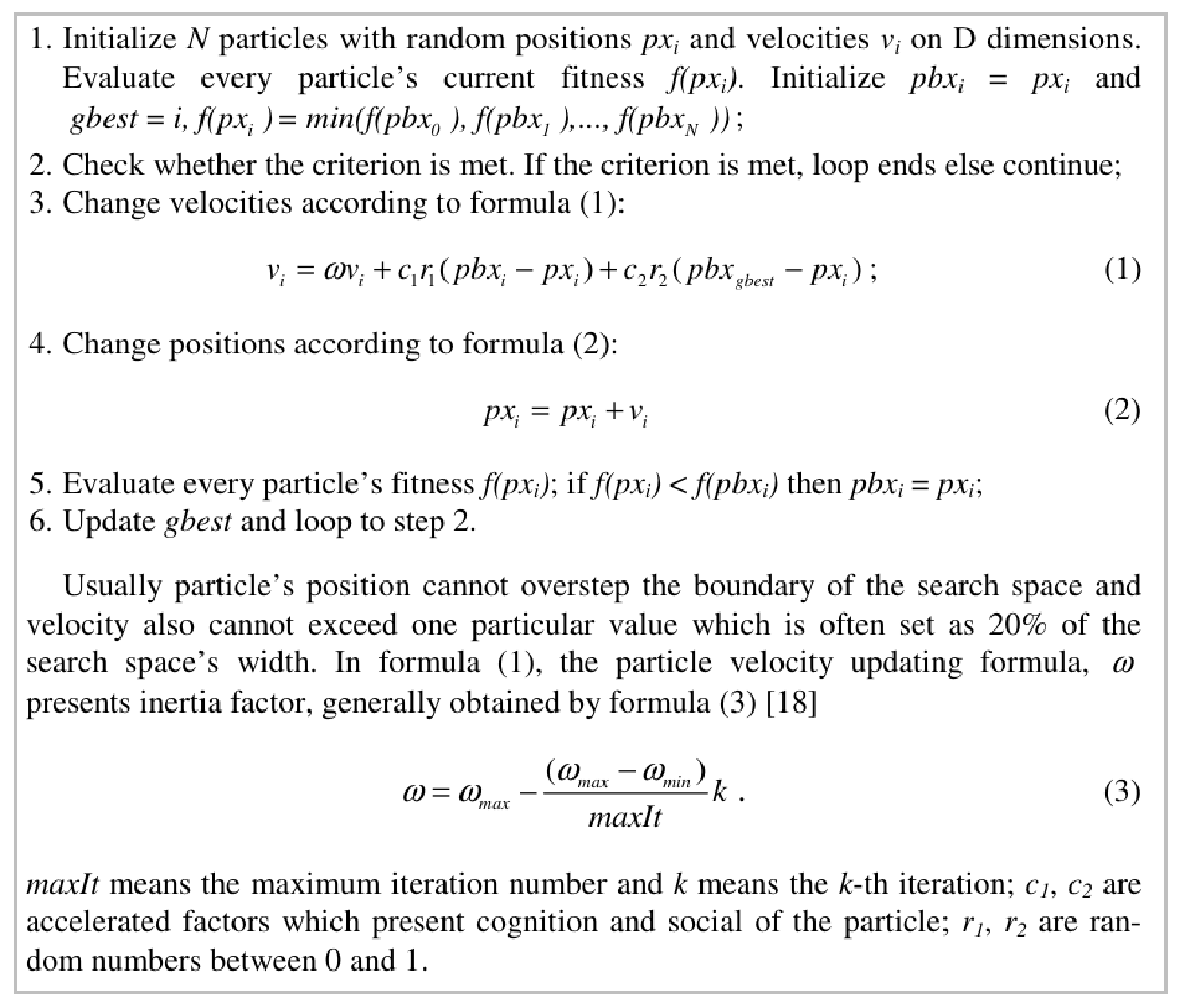}
    \caption{Basic Steps of PSO \cite{Jia-PSO-Test-Generation}}
    \label{fig:PSO}
\end{figure}

In comparison with GA, PSO has been noted to be the superior AUTG approach by multiple researchers. Jatana et al. notes that although GA will match PSO in terms of objectives and search capabilities, PSO on the other hand "requires less computational efforts in comparison to GA" which indicates the efficiency of using PSO in large software code bases \cite{Jatana-PSO-MT}. Nayak and Mohapatra have also computationally compared both GA and PSO. They echo the findings made by Jatana et al. in which that PSO is less computationally "complex". Within GA, the sorting of the population using Quicksort to find the best individual could potentially be a $O(n^2)$ to $O(nlog2n)$ process which can become very computationally expensive the greater the size of \textit{n}. PSO on the other hand, is only required to update the velocity and position of each particle within the population with no need for sorting - resulting in a time complexity of just $O(n)$ \cite{Nayak-PSO-DF}. Windisch et al. concludes that PSO is inexpensive and that it has fast convergence to optimal areas as compared to the slow movements made by GA \cite{Windisch-Apply-PSO}. The analysis above therefore indicates the general acceptance by researchers within the community that PSO is indeed superior over GA. 

We have now explored the two AUTG methods known as Genetic Algorithm as well as the Particle Swarm Optimisation that we had set out within our Scope. The above analysis thus also answers our first research question where we had chosen to better understand how the evolutionary methods operated alongside the added bonus of their differences. In the following section, we will continue to explore the "local minima" problem that the Genetic Algorithm may experience whilst also exploring the various improvements that have been made to both of these AUTG methods. 

\section{Performance Analysis and Improvements} \label{chap:performance and improvements}

In this section we will be reviewing the performance of these automated algorithms through the various experiments conducted by previous researchers. We will also showcase the possible improvements that have been raised by others. These improvements include the approach of creating "hybrid" algorithms where two different evolutionary methods are combined into one to address the shortcomings of the individual methods on their own. These hybrid methods include combining the Genetic Algorithm with other methods such as a Tabu Search. Other improvements include the inclusion of machine learning and neural networks to speed up certain components of the algorithm to obtain a better result. As Genetic Algorithm is the fundamental approach within search-based evolutionary testing, the focus will be upon it. 

\subsection{Random Testing VS Search-Based Testing Performance}

A common point of comparison within AUTG is between Random Testing against the methods of Search-Based Software Testing (SBST). The general understanding is that SBST will be superior in generating more specific and smaller test suites in a shorter period of time in comparasion to Random Testing.

Random Testing and SBST was utilised in an experiment to generate test cases automatically for the Triangle Classification. Triangle Classification (Tri-Typ) is a famous benchmark where its aim is "to classify a Triangle into (1) Scalene, (2) Equilateral, (3) Isosceles, and (4) Invalid Triangle" \cite{Mann-Test-Data-Generation}. Mann et al. have found that for this problem, Random Testing would take around 94 seconds to generate 4,863,962 test cases in order to attain 100\% full path coverage whilst the Genetic Algorithm SBST approach only took 1.88 seconds to generate only 39 test cases \cite{Mann-Test-Data-Generation}. This metric already highlights the weaknesses of Random Testing where it will take significantly more time to generate multiple test cases whilst in stark comparison, SBST only required 39 test cases to obtain full coverage in a significantly reduced amount of time. 

This performance can also be witnessed in the usage of AUTG tools such as Randoop and Evosuite in a financial application. Almasi et al. found that the random nature of Randoop was able to detect 38\% of faults whilst the evolutionary nature of Evosuite was able to detect 56.4\% of faults out of a total 25 real faults \cite{Almasi-Financial}. Almasi et al. also found that search-based methods would be able to improve their coverage and detection rates if the search budget was increased - this resulted in a 5.6\% for Evosuite compared to only 1.2\% for Randoop. The authors conclude that search-based approaches are better due to their innate ability to continuously generate corner cases to detect hard faults which is a capability that Random Testing lacks \cite{Almasi-Financial}. Whilst Almasi et al. were able to achieve higher coverage and fault detection with Evosuite by increasing their search budget, it is interesting to note however that Vogl et al. found the otherwise. Within the SBST 2021 competition, Vogl et al. found that by increasing the search budget from 30 seconds to 120 seconds, an increase in average statement coverage was achieved but overall coverage dropped due to Evosuite "crashing more frequently when given more time" \cite{Vogl-Evosuite-2021}. 

Random Testing however does have some slight advantages. Majma and Babamir found that Random Testing ultimately created too many test cases and thus made it difficult for the developer to separate the useful test cases from the not as useful cases \cite{Majma-NN-Test-Oracle}. They however also confirm that Random Testing would be useful and has adequate performance in smaller programs that are less complex. Random Testing performance will however deteriorate as the software under test grows in size \cite{Majma-NN-Test-Oracle}. Multiple researchers thus conclude the effectiveness and superior performance that Search-based Testing possesses over its Random Testing counterpart. 

\begin{figure}
    \centering
    \includegraphics[width=0.5\linewidth]{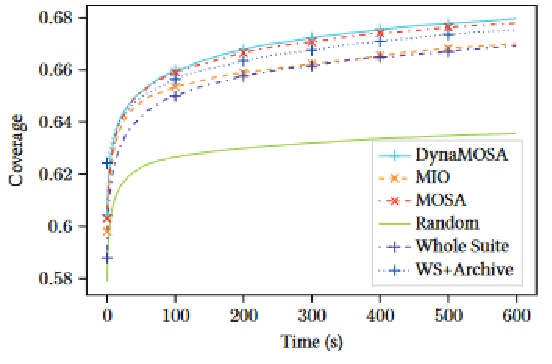}
    \caption{AUTG Coverage Over Time \cite{Lukasczyk-Pynguin}}
    \label{fig:AUTG-Coverage-Over-Time}
\end{figure}

There exists additional research that supports the above sentiment. Within the SBST 2022 competition, Evosuite utilised the Dynamic Many-Objective Sorting Algorithm (DynaMOSA) evolutionary algorithm which is driven by optimised fitness function searches to generate test cases. Schweikl et al. have found that the algorithm utilised a lot of processing both before and after to reduce test flakiness, reduce redundant test cases as well as to also ensure that the test cases are somewhat readable \cite{Schweikl-SBST-2022}. Readability of generated test cases is actually a major challenge within AUTG and will be explored further below. Nevertheless, this approach achieved first place out of the other AUTG methods in the competition \cite{Schweikl-SBST-2022}. This performance can also be seen within another paper where Lukasczyk et al. compared the performance of search-based methods against random testing. As seen in Figure~\ref{fig:AUTG-Coverage-Over-Time}, Lukaczyk et al. found that for the evolutionary search methods including DynaMOSA, the methods would continue to scale and gain increasing coverage as time continued. This was in contrast to Random Testing where the coverage stagnated and stayed mostly the same despite an additional search budget for time. 

Whilst all of these papers showcase the superiority of search-based methods over Random Testing, it is also important to note that these methods also suffer limitations and challenges that are faced by AUTG in general. Lukascyzk et al. found that if the system under test did not possess adequate or explicit type declarations within their functions, AUTG tools such as Pynguin would have to "randomly choose types from all available types, with a low probability of choosing the correct one" \cite{Lukasczyk-Python-Automated}. Type information referred here include strings, integers or custom classes. This is critical as without this type information, the AUTG tools and methods are not able to cover large parts of code that may contain complex objects that are required to be instantiated. This therefore results in the "search" process of the method being crippled down to a random one in all but name. The challenge of statically-typed vs dynamically-typed languages and its implications for AUTG will be explored further below. 

In recognition of the abundance of research supporting search-based testing, we can therefore conclude that search-based testing approaches including their evolutionary approaches within are superior to that of Random Testing. This conclusion thus answers the auxillary question of our first research question where we had set out to better understand the chosen AUTG methods alongside the performance of AUTG methods. In this case, AUTG is indeed superior to that of Random Testing although there are still caveats that needs to be addressed. 

\subsection{Improvements}

In the following section we will explore then various improvements that have been suggested by researchers. These improvements replace or augment certain parts of the original AUTG methods so that performance in coverage can be improved or limitations such as global/minimum search space can be addressed. Most improvements appear to stem from the combination of two approaches into one or by improving the fitness function. There are however, some unique approaches that incorporate the usage of deep learning to improve AUTG coverage and performance. We will now explore them below. 

\subsubsection{Hybrid Algorithm Approaches}
\mbox{}\\
There have been multiple proposal to combine Genetic Algorithm (GA) with Particle Swarm Optimisation (PSO). Within their paper on the usage of GA on smart contracts, Ji et al. found that the previous implementation of "All-uses Data Flow criterion based test-case generation using GA (ADF-GA)" was found to have an unsatisfactory coverage rate of around 77.3\% when generating test cases for the smart contract software. Whilst still superior to Random Testing, Ji et al. argue that the low quality of the test cases generated by the GA AUTG method was in-fact due to "the randomness of the genetic operations" \cite{Ji-DataFlowTesting}. They continued on to claim that when the evolutionary process was executed on top of the low quality genetic operators, the search space would increase alongside the number of iterations it would take to find the optimal solution \cite{Ji-DataFlowTesting}. Further prior research performed by the authors found that previous attempts to combine GA and PSO together was effective. GPSCA is a combined GA and PSO algorithm along with an updated fitness function in which its application resulted in a reduction of overall test cases generated whilst still achieving higher coverage as compared to the individual execution of GA and PSO \cite{Ji-DataFlowTesting}. Based off of the historical validation, Ji et al. thus proposed to utilise PSO to "accelerate" the GA process in finding global optimum solutions. Their approach advocated for the "recombination of parent populations within the evolutionary process" in order to minimise the effect that random genetic operators had at disrupting the search space \cite{Ji-DataFlowTesting}. The parent populations were saved as variables and passed onto the next to guide the population search process. The approach resulted in an improvement in the average coverage rate from the aforementioned 77.3\% to a new average of 89.2\% whilst reducing the amount of time and search budget required to generate the test cases.

\begin{figure}
    \centering
    \includegraphics[width=0.7\linewidth]{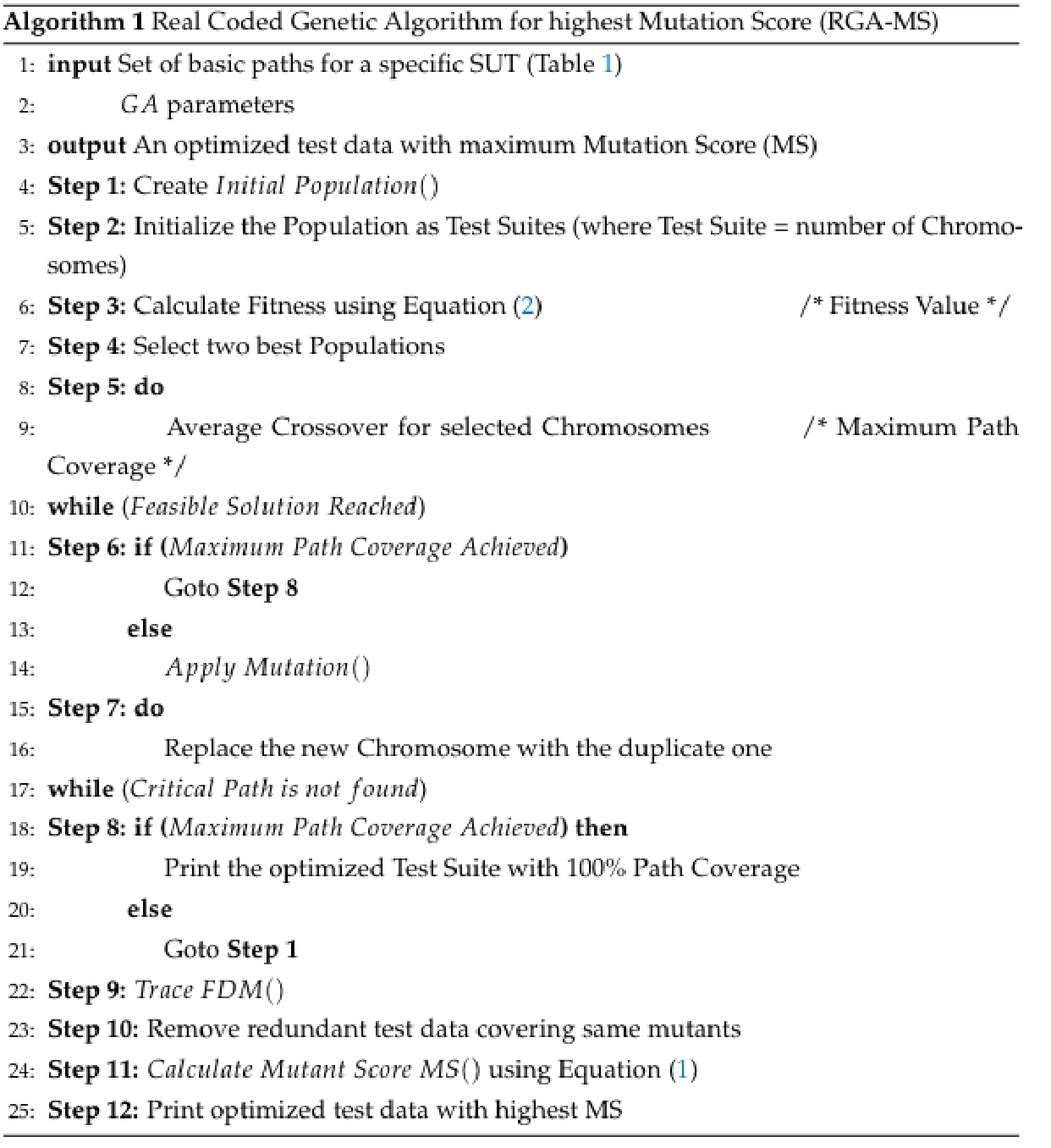}
    \caption{Real Coded Genetic Algorithm for Highest Mutation Score (RGA-MS) \cite{Mishra-Genetic-MT}}
    \label{fig:RGA-MS}
\end{figure}

\begin{figure}
    \centering
    \includegraphics[width=0.75\linewidth]{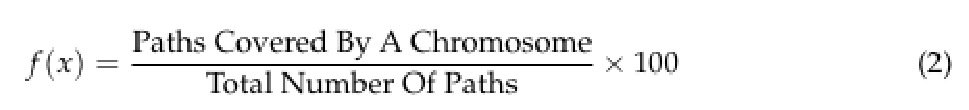}
    \caption{Fitness Function for RGA-MS \cite{Mishra-Genetic-MT}}
    \label{fig:RGA-MS-FF}
\end{figure}

In a continuation of hybrid approaches, there have also been recent attempts at combining evolutionary algorithms with mutation testing in order to achieve better coverage and fault detection performance. As identified earlier, Mutation Testing is the process to artificially emulate faults that a developer may make within the software so that the generated test cases are continuously "evolved" into they can detect the mutated program errors \cite{Jatana-PSO-MT}. The first combined approach incorporating GA with Mutation Testing was introduced by May et al. back in 2007. The hybrid approach first commences by randomly initiating a starting population of initial test cases with each test case represented as a chromosome. The fitness function of the GA will subsequently apply the standard selection, crossover and mutation process to determine the next generation of test cases. They key distinction here however is that the fitness function is guided by the mutation score of the test cases based on how many mutants they can eliminate within the mutated software \cite{Jatana-PSO-GA-MT}. Mishra et al. builds upon this hybrid approach by proposing their "Real Coded Genetic Algorithm for Highest Mutation Score (RGA-MS)". RGA-MS incorporates the usage of a "Fault Detection Matrix (FDM)" alongside optimal path coverage in order to achieve maximum mutation coverage \cite{Mishra-Genetic-MT}. The authors state that the integration is necessary as current path coverage-based testing methods may "fail to detect all software faults because the information about fault detection is not incorporated into the process of generating test data" \cite{Mishra-Genetic-MT}. The algorithm for RGA-MS can be viewed in Figure~\ref{fig:RGA-MS} where equation 2 refers to the Fitness function used to find the optimal test cases defined in Figure~\ref{fig:RGA-MS-FF}.

RGA-MS follows the standard mutation process before it arrives at the FDM step. The purpose of the FDM and the RGA-MS approach is to minimise the amount of test cases required in the test suite. The FDM will rank test cases by mutation score and will retain the smallest amount of test cases that can cover all of the identified mutants within the code \cite{Mishra-Genetic-MT}. Mishra et al. conclude that whilst this approach is computationally expensive due to the Mutation Testing process, it will result in a smaller amount of test cases which will help developer workflows and readability. The approach also ensures higher quality test suites as well as reliability since the method would have generated a optimal test suite that can eliminate all of the identified mutant faults within this software \cite{Mishra-Genetic-MT}. Whilst the hybrid approach is a good improvement, other researchers including Mishra et al. themselves have identified that GA-MT is inferior to the hybrid combination of PSO and Mutation Testing below.  

\begin{figure}
    \centering
    \includegraphics[width=0.8\linewidth]{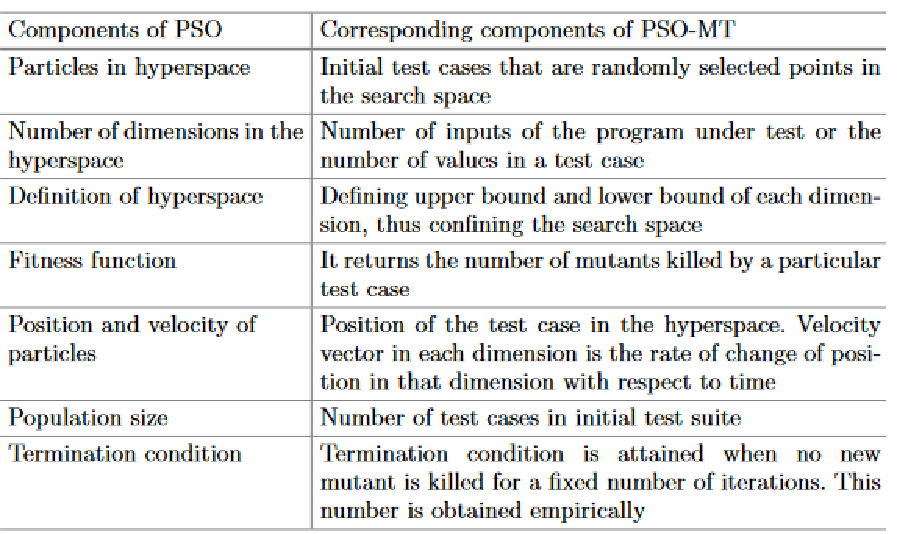}
    \caption{Mapping Components of PSO to PSO-MT \cite{Jatana-PSO-MT}}
    \label{fig:PSO-MT-Components}
\end{figure}

The next Evolutionary-Mutation hybrid of interest is the combination of PSO with Mutation Testing. Within their paper outlining the proposal of RGA-MS, Mishra et al. also identified that a PSO-MT equivalent would in fact be superior to their Genetic Algorithm counterparts specifically in terms of computational efficiency \cite{Mishra-Genetic-MT}. Prior research by Hassan et al. found that although GA-MT hybrids had the same effectiveness as PSO-MT hybrids, PSO implementations had better computational efficiency over their GA counterparts which echos the earlier findings that were made by Mishra et al \cite{Jatana-PSO-MT}. Within the same paper. Additional historical context from Li et al. found that GA possessed "high global search ability with low computational efficiency, low speed as well as a difficult/slow convergence" which thus backups the notion where GA-based methods are inferior in nature to PSO-based methods \cite{Jatana-PSO-MT}. In a later paper that compared the performance of GA-MT with PSO-MT, Jatana et al. found that in a scenario with infinite time, both approaches would deliver a complete search within the program and kill off the mutants. The paper also found that less test cases were generated in a PSO-MT approach due to "PSO-MT converging faster than GA-MT in most cases". This was as a result of the PSO AUTG method only requiring a smaller amount of particles to kill identified mutants which contrasts the higher number of chromosomes required by their GA counterpart \cite{Jatana-PSO-GA-MT}. In order to match the PSO method to its new combination with Mutation Testing, its components was updated in order to match the new "definitions" as seen in Figure~\ref{fig:PSO-MT-Components} that are in line with the mutation process. 

\begin{figure}
    \centering
    \includegraphics[width=0.55\linewidth]{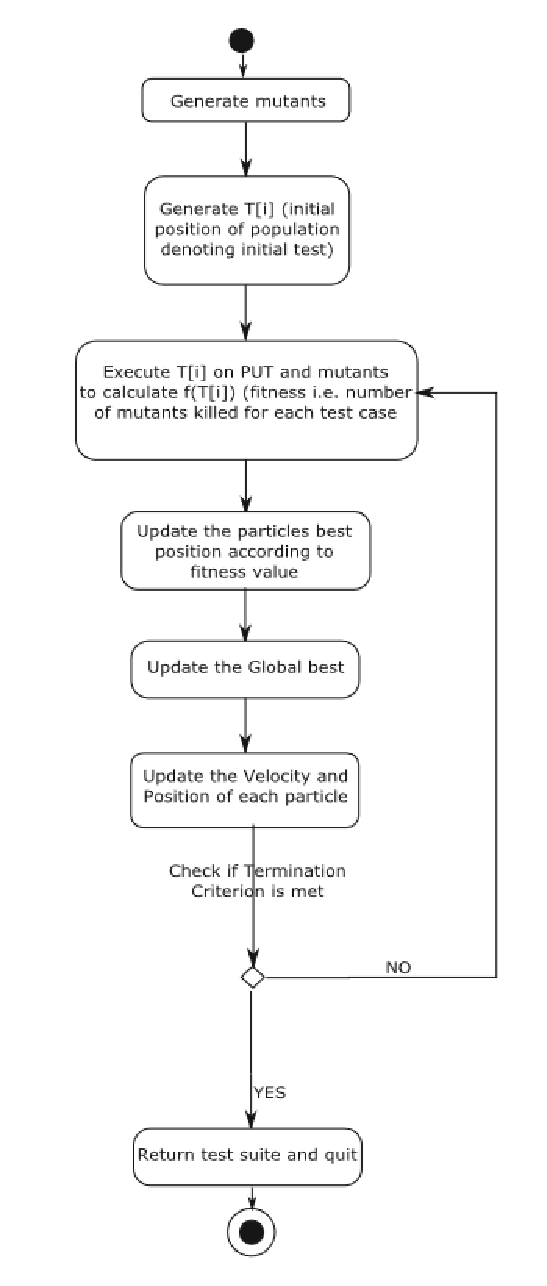}
    \caption{PSO-MT Algorithm Flow \cite{Jatana-PSO-MT}}
    \label{fig:PSO-MT}
\end{figure}

Within the PSO-MT approach proposed by Jatana et al., mutants and points within the search space that represent the test cases are randomly generated. The test cases are then executed on the mutants to which its performance is subsequently evaluated by a fitness function to determine the effectiveness of the test cases. The velocity and position of the particles within the search space are then updated to continue finding the optimal test suite. The whole process continues until the termination criteria is reached and the algorithm terminates \cite{Jatana-PSO-MT}. The flow of the algorithm can be seen in Figure~\ref{fig:PSO-MT}.

As identified by Jatana et al. as well as Mishra et al., Genetic Algorithm has been found to have a slow convergence. Mayan and Ravi discuss this tendency within their paper and raise that the GA suffers from issues such as "slow convergence, blind search and the risk of being stuck within the local optimum solution" \cite{Mayan-Hybrid-Optimisation}. This is important as most meta-heuristic search methods actually view the testing problem as a search space optimisation problem. As a result of the slow/premature convergence, AUTG search methods such as GA may not actually be satisfying the requirements of this optimisation problem as they present a solution that may not actually be the global optimum due to being stuck elsewhere in the search space. This issue pertaining to GA was addressed above as we explored methods of combining GA with mutation testing in order to better the guide the evolutionary process. 

Mayan and Ravi argue for the usage of the Tabu Search algorithm to be incorporated within the GA process. Tabu Search is a memory-based algorithm that "guides local heuristic search procedure to explore solution space beyond the local optimally" and keeps track of all of the paths that have already been explored in memory so that redundant searches aren't made again \cite{Rathore-GA-Tabu}. The authors argue that the addition of this algorithm to save GA-generated "good" candidates within the Tabu Search to be used again during the mutation step would reduce execution time and randomness of the original mutation process \cite{Mayan-Hybrid-Optimisation}. This is as a result of the backtracking process within the Tabu Search algorithm that allows the search to move away from the local optima \cite{Rathore-GA-Tabu}. Unlike the original randomised mutation process, the Tabu Search look-ups would better guide the evolutionary process to select the "minimum set of test cases based on their high code coverage, less execution time and high fault detection rate". In another similar attempt, Rathore et al. advocated the use of two Tabu lists instead of just one as used by Mayan and Ravi. The first Tabu list is known as the "Short-Term Tabu List" which would be populated by inputs that are commonly known to have low fitness predicates. The second list is known as the "Long-term Tabu List" and would be the list used in the backtracking process. Rathore et al. state that the usage of the two Tabu lists would ensure that low fitness inputs are avoided whilst the highest fitness predicates of each iteration were kept and used as a guide \cite{Rathore-GA-Tabu}. Whilst out of scope for this paper, another approach involving the addition of Tabu Search was made by Srivastava et al. where they combined Tabu search with Cuckoo search to form CSTS. The authors of that algorithm argued that Cuckoo search shared similar limitations to GA where the algorithm would also converge towards the local optima \cite{Srivastava-CSTS}. Cuckoo search is however superior requiring less variables than GA whilst also requiring minimal time to converge towards a solution. The Tabu search component would help the overall algorithm backtrack if it fell within a local optimum \cite{Srivastava-CSTS}. The inclusion of Tabu search within AUTG methods thus resulted in more directed and effective test suites that were more "optimal" as compared to their unary and uncombined counterparts.

\subsubsection{Genetic Algorithm - Overcoming the Local Optimal Solution}
\mbox{}\\
As we identified earlier in statements made by Jatana et al. as well as Mishra et al., the Genetic Algorithm is known to have a slow convergence as well as the risk of being stuck in the local optimum solution. Similar observations have also been made by other researchers. A point of interest that could be made would be to have smaller population sizes so that the local optimum is not too far off the global optimum. Mann et al. however disagrees and states that "although smaller population sizes finds quick solutions, their premature convergence rate results in sub-optimal solutions" \cite{Mann-Test-Data-Generation}. The authors continue to state that the smaller population size would result in lower diversification which in turns prohibits goal search. A larger population size, albeit with a higher processing cost as compared with Random Testing, would allow for "increased diversification that has fewer chance to be stuck in the sub-optimal area" \cite{Mann-Test-Data-Generation}. Despite the increase in population size to diversify, the "exploration ability" is still affected by mutation operators as well as the crossover step \cite{Kifetew-Orthogonal}. The increase in population however, is not beneficial if the individuals that make up that population are not diverse in nature. Kifetew et al. presents the phenomenon present within Genetic Algorithms known as "Genetic Drift". Genetic Drift inhibits the ability of evolutionary algorithms to diversify their search (despite a large population) in search of other potential solutions as the search is dominated by a small set of individuals that are too similar to each other \cite{Kifetew-Orthogonal}. This leads to an premature convergence to a sub-optimal solution that may represent a local optimal solution rather than the true global solution. This view has also been supported by Hu et al. to which they state that "if the population diversity is too low, the evolution of ET may be trapped into a local optimal solution" \cite{Hu-Hybrid-Optimisation}. Hu et al. however also warn that if the diversity of the population is too high, then the individuals will become to decentralised to converge which makes the entire AUTG approach no better than Random Testing \cite{Hu-Hybrid-Optimisation}. There have also been other papers that do not support the view on local optimum. Although the authors don't specifically present a case on this matter, Rani et al. states that within the solution generation process, evolutionary algorithms will perform two operations which are "Intensification" and "Diversification". The "Intensification" process will "search the neighbourhood search space and exploit the solution by selecting the best of the local solutions" whilst the "Diversification" process will "explore the search space globally and try to diverse the solution" \cite{Rani-Elitist-GA}. The ability to either search locally or diverse from the global search indicates that evolutionary algorithms are able to "escape" from the local optimum solution which is a view that is shared by Derderian et al. Within their paper, Derderian et al. have stated that GAs are particular useful since "one of their benefits is the ability to escape the local minima in the search for the global minimum" \cite{Derderian-Temporal}. This view has also been echoed by Liu et al. to which they claim that "Compared with other groups of intelligent optimisation algorithms, Genetic algorithm has good global search ability" \cite{Liu-RBF-NN}. These statements thus appears to be contradictory to the earlier views established by earlier researchers above and thus indicates this to be an issue of contention.  

Regardless of the viewpoint presented, the researchers above have made attempts at addressing and improving this issue. As the issue of local optimums relates to diversity within the population, the attempts have been made at improving that respective diversity. Li et al. presents a hybrid algorithm combination of GA alongside Simulated Annealing (SA). They convey that the "probability of mutation is no longer a constant" within hybrid GA and so that the mutation process won't affect diversity as much the closer the process gets towards the global optimum - Allowing the authors to change the mutation probability dynamically \cite{Li-Configuration}. Kifetew et al. initially agrees with this statement as they also announce that a simple technique for introducing diversity is to modify the mutation rate as a "higher mutation rate clearly increases the ability to explore more areas" \cite{Kifetew-Orthogonal}. The authors however also conclude that this process would also "prevent the convergence towards any optimum solution" and thus reduces the ability to locally search for any better solutions \cite{Kifetew-Orthogonal}. They also remark that this approach at improving GA would not "guarantee the exploration of un-explored spaces within the population and may also result in spaces being re-explored" \cite{Kifetew-Orthogonal}. Kifetew et al. are also concerned that even if diversity was introduced within the population to address the issue of local optimum solutions, the "issue of generating individuals (solutions) already considered in the past would still remain \cite{Kifetew-Orthogonal}. 

\begin{figure}
    \centering
    \includegraphics[width=0.6\linewidth]{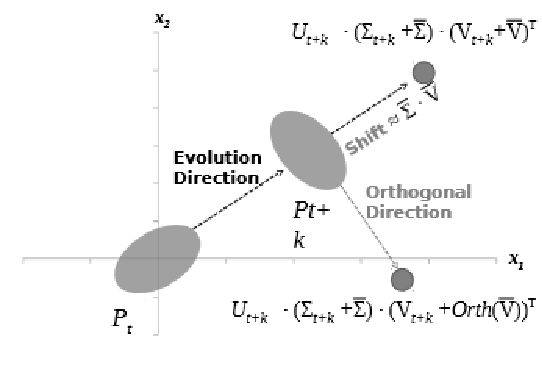}
    \caption{Graphical Interpretation of Orthogonal Diversification - SVD-GA \cite{Kifetew-Orthogonal}}
    \label{fig:SVDGA}
\end{figure}

The issue of solutions generated that have already been considered has been addressed by Rani et al. through their implementation at an elitist GA approach. Within a general implementation of the Genetic Algorithm, Rani et al. states that the "individuals of a current population are removed and new individuals are derived using reproduction over the current population" \cite{Rani-Elitist-GA}. This indicates that the population is continuously regenerated with new iteration which means that the "best" individuals of a particular generation are not being saved and thus goes against the principle of "survival of the fittest" within biology. As a test case will retain the same "fitness" within the entire test, Rani et al. reflects that it is best to be "elite" and keep the "fittest" test cases so that extra time and resources are not spent regenerating a similar test case that was already previously created \cite{Rani-Elitist-GA}. This approach is similar to the Cataclysm algorithm that Hu et al. have proposed in light of a premature and non-diverse population. Hu et al. presented the case where if the diversity of a population was too low that a "cataclysm operation" should occur where the "best" individuals within a population are kept and a new population is regenerated and merged with the previous "best' individuals - similar to the elitist GA algorithm process \cite{Hu-Hybrid-Optimisation}. Kifetew et al. also provides a novel solution in which they expand the search space in an "orthogonal" direction as seen in Figure~\ref{fig:SVDGA} in order to more effectively explore the search space and escape the local minimum solution. The approach consists of three parts which includes: "History Aware Orthogonal Exploration"; "Reactive Exploration"; as well as "Reactive Orthogonal Exploration". Kifetew et al. explains that the population history is considered as part of the "History Aware" process. They also state that random directions are also incorporated when the orthogonal direction yields no improvement and "null evolution direction" occurs. These two approaches are also combined to form the "Reactive Orthogonal Exploration" process as part of the algorithm approach \cite{Kifetew-Orthogonal}. 

We can therefore see that this is a major issue of contention present within the Genetic Algorithm. For our review, there seems to be a group of researchers that believe that GA has a "Genetic Drift" problem where it will become stuck within the local optima whilst other researchers have presented views that go against this view. Some researchers have also implemented suggestions at overcoming the local optimum issue through the usage of elitist algorithms and/or by increasing population diversity. Nevertheless, it would appear that further research into this issue is necessary. 

\subsubsection{Neural Networks and Generative Adversarial Networks}
\mbox{}\\
Researchers have spent a lot of time in improving the respective fitness functions in evolutionary AUTG methods such as the Genetic Algorithm as seen above. Apart from research seeking to improve the various parameters within the function, there has also been research conducted into replacing and substituting the fitness function step through the usage of Deep Learning and Neural Networks. We will now proceed to explore those contributions below. 

Pan et al. introduces a new approach at improving GA by simulating the fitness function step with a Back-Propagation Neural Network. The authors presented the case that within the original GA method, the process of using the fitness function to determine the probability of the individual within the population being selected for inheritance "wasted a lot of time" and thus was not computationally efficient \cite{Pan-Method-Of-Generating}. Liu et al. have also been in agreement where they state that the process of calculating the fitness values as part of the fitness function would take a lot of time and work which would "seriously affect the speed of test case generation" \cite{Liu-RBF-NN}. In light of this, Pan et al. utilises the trained neural network to predict the fitness value of the population during the run-time of the GA process based on path-coverage. The accuracy of the deep neural network has been identified to be affected by the following factors: the depth of the neural network; the number of the nodes in the hidden layer; as well as the activation function of the neurons \cite{Pan-Method-Of-Generating}. Pan et al. warn that if the number of nodes within the neural network is too low then the problem cannot be fully represented by the DNN whereas if the number of nodes is too high then it will be over-fitting and will not be able to generalised future data \cite{Pan-Method-Of-Generating}. Nevertheless, the approach has been found to be superior than the original GA process and significantly uses less time as the program under test grows in complexity \cite{Pan-Method-Of-Generating}. Another approach at a DNN was made by Liu et al. where they introduced a combined GA-RBF (GAR) neural network approach in order to "simulate the fitness function of the branch coverage path to solve the fitness value of the population" \cite{Liu-RBF-NN}. Unlike the path-coverage approach made by Pan et al., this new approach utilises branch-coverage which we identified to be more strict and one of the strongest coverage criteria. The DNN approaches made by both parties all aimed to address the issues of premature and slow convergence that GA exhibit \cite{Liu-RBF-NN} \cite{Pan-Method-Of-Generating}.  

Research has also been done in the usage of Generative Adversarial Networks (GAN) within the AUTG method space in hopes of potential improvements to the original SBST methods. Guo et al. have defined GANs to be "algorithmic architectures that use two neural networks" in order to "pit" them against each other (hence adversarial) in "order to learn the underlying distribution of the training data so that it can generate new data instances" \cite{Guo-GAN}. A GAN also consists of a generator and a discriminator in which their purpose is to "reinforce the ability of data generation from the generator by competing to the discriminator to detect fake data" \cite{Guo-GAN}. In the approach proposed by Guo et al., the GAN will generate test inputs and paths from the generator whilst the discriminator will "learn" the execution paths. The generator will then generate test cases in order to match the discriminator. Performance results from within the paper indicate that the GAN approach will outperform Random Testing in terms of 100\% branch coverage. However, Guo et al. also identified that the GAN approach consumed more time compared to Random Testing and that the training process also took a lot of time. In light of this performance, it is possible to conclude that GANs do not pose as a potential improvement for evolutionary algorithms such as GA and that individuals should continue to focus on neural networks instead to improve the mutation process. This conclusion drawn by us however, is countered by Zhang et al. to which they state that existing DNN approaches are not effective. The authors also state that a large number of redundant test cases are generated of which "do not meet the test requirements or the actual situation" \cite{Zhang-Condition-GAN}. It is thus difficult to conclude the benefits that the usage of neural networks have on AUTG methods and we therefore believe that further research within this space is necessary. 

In a continuation from above, multiple researchers including Zhang et al. have identified a need for better testing of neural networks through the usage of novel test criterion specifically used for neural networks in order to ensure reliability and accuracy of the generated test cases. Zhang et al. proposes a coverage-guided GAN known as CAGTest. CAGTest is able to determine the potential faults that may occur from other DNNs during the test generation process to which it will itself generate test cases conditionally to identify them in an efficient matter \cite{Zhang-Condition-GAN}.

\subsubsection{Other Improvements}
\mbox{}\\
Whilst not the primary topic of this paper, we note that there have also been improvements made towards Random Testing. An interesting case that we would like to explore is the combination of Random Testing with the evolutionary approaches such as the Genetic Algorithm proposed by Chen et al. Adaptive Random Testing (ART) was first proposed by Chen et al in an effort to increase the fault-detecting abilities of the original random approach. This approach however was found to "only increase the distance between the test cases to enhance its failure detection ability". This was as a result of the fact that inputs that caused failures in the first place were generally "clustered" together \cite{Mao-ARTMoes}. It was thus in the best interest of the developer to "space" out and have more "diverse test cases within the program to detect these clustered errors. Instead of simply increasing the space around in search of the next test case, Tappenden and Millar proposed the combination of ART and evolutionary methods such as GA to convert the sampling process into a global search problem which would be handled by the GA \cite{Mao-ARTMoes}. This approach was subsequently known as eAR and had high effectiveness in finding faults within the program under test. 

\begin{figure}
    \centering
    \includegraphics[width=1\linewidth]{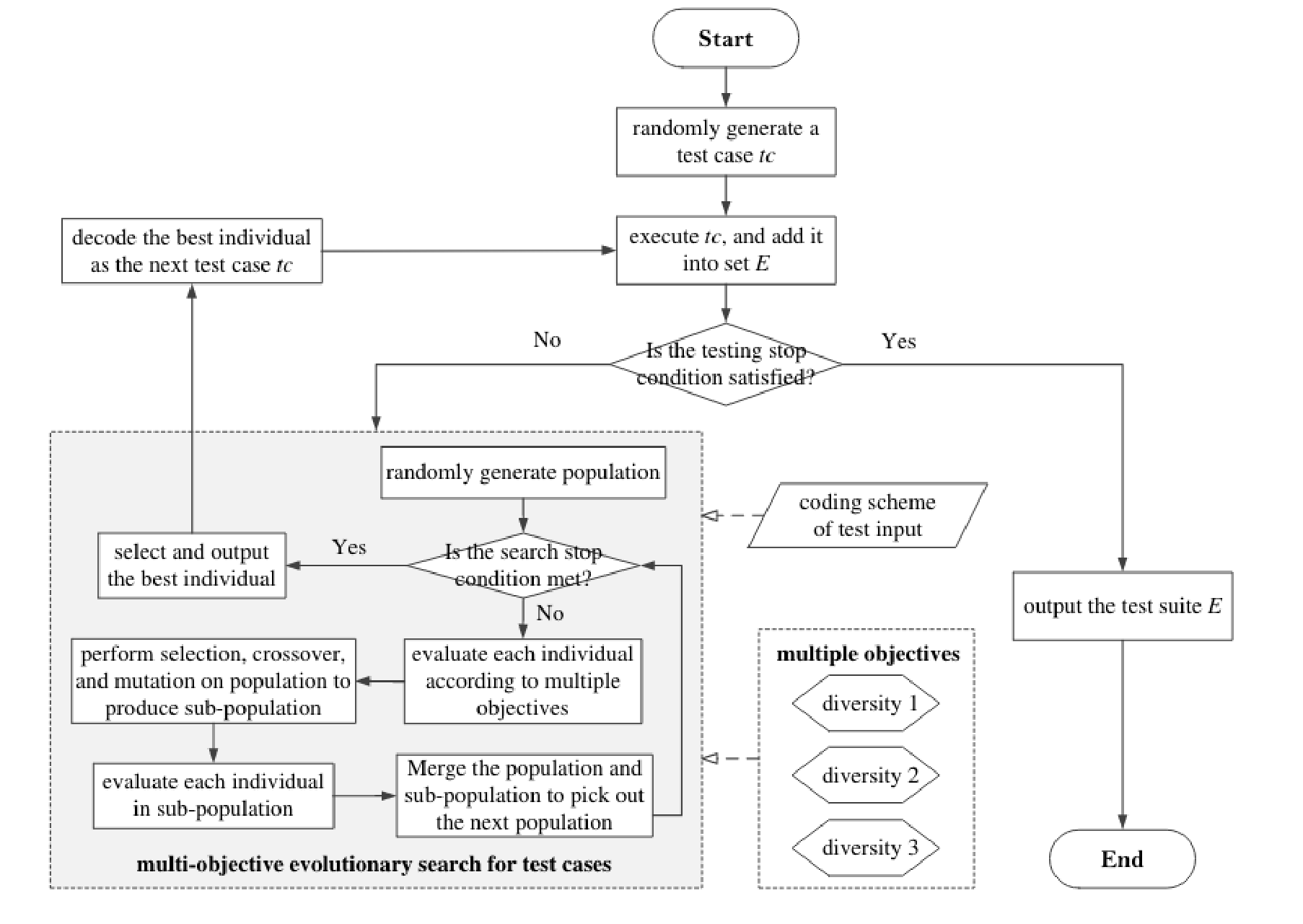}
    \caption{Multi-Objective Evolutionary Search-Based ART \cite{Mao-ARTMoes}}
    \label{fig:MoesART}
\end{figure}

Mao et al. improved upon this approach by proposed an evolutionary approach that was multi-objective in its search, forming MoesART. The authors argued that the previous eAR approach was only considering the diversity of the test cases from one perspective which may not lead to a uniform distribution of test cases \cite {Mao-ARTMoes}. They thus proposed that each individual generated within the population would be evaluated from three different perspectives of diversity which were: dispersion; balance; as well as proportionality. "Dispersion" refers to the spreading of generated test cases as widely as possible within the input domain so that it shouldn't too close to an already present test case. "Balance" however states although it is good to keep all of the test cases as far away from each other, it is important to ensure that they are all still "uniform" and are of equal distance from their neighbours. The final objective of "Proportionality" stated that within the different faults identified within the program, the amount of test cases for a particular fault should be proportional to its size and weight of importance \cite{Mao-ARTMoes}. It is not ideal to have a small fault have multiple test cases all testing for the same fault but a major fault not having diversity in its test cases. The diversification of the population also addresses the issue of local optimum solutions that we explored earlier above. The above objectives thus forms the "multi-objective" part of the search which can be seen in Figure~\ref{fig:MoesART} during the method process. 

After applying the MoesART method in practice, Mao et al. concluded that the algorithm had "better failure detection ability than the single-objective evolutionary search-based ART" which was also known as eAR \cite{Mao-ARTMoes}. The authors also found that the failure detection was significantly enhanced in two and three-dimensional input domains \cite{Mao-ARTMoes}. MoesART and eAR thus formed examples of cases where the original underlying random testing approach was combined with evolutionary methods to improve their fault detection abilities within a system under test. 

Another attempt at improvement was presented by Bader et al. to which they execute the Genetic Algorithm on multiple CPU threads in an attempt to "parallelize" the algorithm. Bader et al. have stated that the Genetic Algorithm has been successful in generating unit tests but also takes too long to produce reasonable results and has large variance \cite{Bader-Parallel}. They thus propose that the search process be "divided over a multi-core CPU with each 'strand' or 'process' being able to evolve and occasionally share information with each other" \cite{Bader-Parallel}. The authors however warn that the sharing of results needs to be done carefully and is balanced to avoid premature convergence with the population being extended and built upon rather than leading to identical population models. This attempt by Bader et al. has been shown to be successful with an average code coverage of 84\% on eight sub-populations as compared to the 79\% coverage running the method on a single client or thread \cite{Bader-Parallel}.

In conclusion, the section thus showcases the various improvements that have been made to both GA and PSO. GA and PSO have at times been combined to form a hybrid algorithm whilst in other cases, have been combined with a Mutation Testing variant. Other improvements include addressing the concerns of the local optimum minimum problem that is a current issue of contention between researchers for GA. Finally, the usage of neural networks to act as a "step-in" for fitness calculations have also been utilised. This improvements and their extensive analysis above thus forms the answer for our second research question where we set out to explore the various improvements that have been made to evolutionary algorithms. Within the next section, we will answer our final research questions and present to the user the main reasons why AUTG still has not been adopted en masse and explore the various challenges and limitations that AUTG currently faces. 

\section{Challenges and Limitations} 
\label{chap:challenges and limitations}

Automatic Unit Test Case Generation (AUTG) as a general field of study has had a plethora of research built up across the previous few years. Earlier, this paper discussed the major search-based evolutionary algorithm approaches that exist within AUTG as well as the various improvements in performance and functionality that other researchers have found to augment the performance of the original methods. Researchers have also devised courses of actions to address certain limitations that were prevalent throughout those algorithms such as the local minimum search space problem. Despite these improvements however, the field of AUTG in general suffers from major issues that threaten its widespread adoption and application. The issues are widespread spanning from poor readability of generated unit tests, the method's lack of understanding for complex object classes that have special requirements, as well as the inability to interact with the Document Object Model (DOM) of web applications. We will now explore these issues below. 

\subsection{Environmental Interactions \& Functional Mocking}

A prevalent challenge that faces the field of AUTG today is the factoring of external I/O operations that need to be performed as part of a code function within a test case. This concept is known as "Functional Mocking" and is where the test case will attempt to mock out an external operation or resource in order to simulate its behaviour with the code currently under test. Functional Mocking goes hand in hand with interactions within the environment as a mocked object is generally an I/O operation that cannot be executed in a test environment. Within their paper discussing the present challenges, Panichella identified 2 out of 4 open challenges related to mocking. The first challenge described the need for some functions to perform database operations in order to store and retrieve data. The other challenge referred to how some functions were expecting external data sources such as XML or JSON files. Whilst AUTG tools such as Evosuite can use functional mocks to mimic the calls to external files, "some classes under test might need files with specific content to satisfy some branch conditions" \cite{Panichella-Beyond-Unit}. Panichella also states that reaching 100\% of coverage for a function requiring database access would require a database to be initialised and corresponding code to be generated for both the SQL code and the original function code which very much represents an open challenge to this day. \cite{Panichella-Beyond-Unit}. Evidence of these challenges can be seen in Fraser and Arcuri's application of Evosuite to 100 Java SourceForge projects, where the authors have found that "71\% of all classes lead to some kind of FilePermission error" which indicates that 3/4 of all the classes within the software under test attempted to interact with external files \cite{Fraser-Sound-Empirical}. 

Fraser and Arcuri have stated that "Real-world software often interacts with its environment". The authors build upon this concept and expand on the notion presented by Panichella earlier by stating that a system under test can "read and write files, open TCP connections with remote hosts, start GUI windows, react to mouse/keyboard events" and much more. These wide range of possible environment interactions thus indicate the need for AUTG methods to properly test these external interactions and to not simply mock them away. Fraser and Arcuri comment that the failure to test these environmental interactions properly could result with inputs that deletes the entire file system and more \cite{Fraser-Evosuite-Challenges}. The environmental problem is a current challenge for the Evosuite tool where running it on classes with I/O operations would result in its search leading to the creation of potentially thousands of files within minutes. \cite{Fraser-Evosuite-Challenges}. In their paper identifying the various challenges that currently face Evosuite in real code deployments, Fraser and Arcuri identify that their tool still lacks the necessary features to properly account for real code environments and also conclude that inability to properly account for environmental dependencies is the reasoning for poor coverage.

In another paper on the "soundness" of testing, Fraser and Arcuri presents the first empirical study where they analyse the performance of their tool Evosuite against a wide range of open-source projects that were chosen in a systematic way without bias. The authors highlight the issue where software testing case studies are not chosen in a systematic way which threatens the validity of that study. They state that this is common for industry research as obtaining the data is a "very difficult and time consuming activity" and so there may be no real choice by the researchers \cite{Fraser-Sound-Empirical}. However, Fraser and Arcuri have found that for open-source projects, researchers are still not conducting case studies in a systematic manner. Certain software artefacts that utilised I/O operations would be purposely omitted if it was not supported by the testing method \cite{Fraser-Sound-Empirical}. This sentiment is also echoed by Ramler et al. where the state that there is still a gap on the usage of AUTG techniques proposed by research on actual software \cite{Ramler-AUTG-Retrospective}. Ramler et al. comments on how AUTG techniques are predominately focused on a research setting that is not reflective of real applications.  Interestingly, the findings by Arcuri and Fraser echo the similar points that was presented in the systematic literature review into Search-Based Testing raised by Ali et al. As we discussed earlier in the paper in our related work analysis, Ali et al. questioned the legitimacy and reported that the papers under review were "non-credible" as they were not systematically comparing performance of the SBST algorithms. In a similar case here, the case studies performed aren't a true representation due to the avoidance of I/O operations despite the I/O operations being a very common part of software today. This thus results in case studies not truly accounting for the environment of the AUTG method in their respective analysis. 

\begin{figure}
    \centering
    \includegraphics[width=0.6\linewidth]{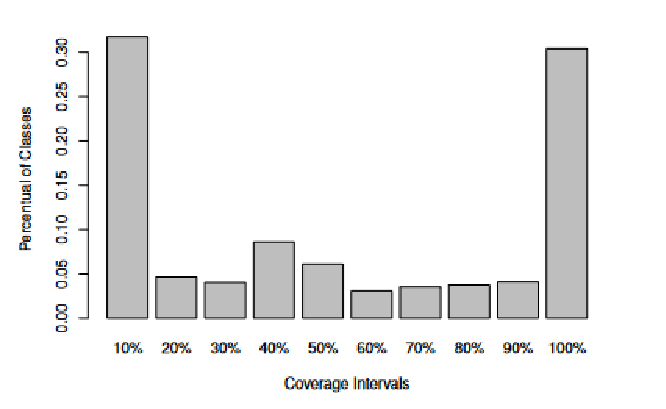}
    \caption{Coverage Intervals of the Classes Tested \cite{Fraser-Sound-Empirical}}
    \label{fig:Fraser-Coverage}
\end{figure}

As a result of their case study, Fraser and Arcuri also raise how permissions can directly affect the environment and thus the generation and execution of test cases within a system under test. Within their deployment of Evosuite into 100 different Java projects, they have found that a wide majority of classes that reported 100\% code coverage which suggested that Evosuite was easily able to account for them \cite{Fraser-Sound-Empirical}. However as seen in Figure~\ref{fig:Fraser-Coverage}, there is also a large number of classes that have very low coverage to which the authors have concluded is as a result of certain permissions not being present and met during runtime. 

The dominant permissions identified include "RuntimePermissions" and "NetPermissions" which accounted for 52\% and 49\% of all the classes within the 100 projects respectively. The two permissions involve actions related to the environment as well as interactions with the network. The classes under test that were checking for "RuntimePermissions" ended up being related to platform-dependent libraries that were loaded for GUI purposes. Each library loaded would in effect trigger a permission check to see if could be executed. On the other hand, "NetPermissions" were common as most of the applications under test were in fact web applications and so the "NetPermissions" would load on any call related to the network such as invalid URLs and other parameters \cite{Fraser-Sound-Empirical}. Having these permissions set properly and remain the same throughout the test generation process is also vital. In their attempt to understand just how effective AUTG tools such as Randoop and Evosuite were at identifying real faults, Shamshiri et al. found that 15.2\% of all of the generated test cases for the Defects4J dataset were flaky. This meant that their "passing/failing behaviour was temporary due to environmental and other dependencies" and thus rendered them useless \cite{Shamshiri-AUTG-RealFaults}. As a result of this analysis, we can see how multiple classes all interact with their environment in some way or form and are "a prime source of problems in achieving coverage". This therefore signifies the importance of having the correct environment for AUTG tools to achieve high coverage which is still a significant challenge and research focus today.

As Fraser and Arcuri have concluded earlier, an open environment that is properly configured can lead to high coverage. Apart from the external environment however, a present problem that is also being faced is the environment within where AUTG methods are presently unable to access internal and private function methods. Shamshiri et al. have found that the AUTG tools struggle to achieve decent code coverage within classes that possess a high amount of private methods that cannot be accessed. In circumstances like this, manually written tests by the developer ended up with a higher coverage than what the AUTG tools could achieve \cite{Shamshiri-AUTG-RealFaults}. Private methods remains a issue and "presents an additional challenge for AUTG, which usually only tests the public interface of a class" \cite{Shamshiri-AUTG-RealFaults}. Other AUTG methods also possess limitations when attempting to account for the environment. In their paper proposing automatic unit testing within embedded applications, Zhang et al. notes that the AUTG method of Dynamic Symbolic Execution (DSE) is unable to properly symbolise environmental variables. The authors note that environmental calls and I/O operations are difficult to symbolise due to its limitation on understanding the semantics of the actual functions themselves which results in a broken symbolisation and low coverage \cite{Zhang-Smartunit}. The AUTG method also has issues mocking pointers properly and requires an "infinite" array to keep track and symbolise all of the pointer operations that a class may perform within embedded applications which ultimately results in more memory used \cite{Zhang-Smartunit}. Incidentally, another similar internal environmental problem was raised by Riberio et al where they announce that one of the challenging issues is related to the "state" problem. The authors explain that this issue occurs when methods exhibit state-like qualities "by storing information in internal variables that are protected from external manipulation" \cite{Riberio-Strategy-Evolutionary}. This inhibits the ability of AUTG tools for testing as the state becomes a black-box state where the only way to observe or modify the "state" is through the execution of method call operations that directly target that "state". Variables possessing the "state" problem alongside internal private methods of classes under test all represent the present environmental issue that is currently a challenge alongside their external counterparts. 

Setting the environment aside and returning to the issue of functional mocking at hand, Arcuri et al. reflect on the need to mock objects that have to be instantiated and configured properly but are otherwise unavailable during a testing environment. In one of the code examples provided by the authors, the execution of a method was performed by passing in a root interface as part of the parameters. This method would have never been called by the AUTG tool due to the absence of a "root" user. This however is readily solved through the usage of a mocking framework such as Mockito. The issue of private methods not being able to be accessed by AUTG as raised earlier by Shamshiri et al. can also be solved through the usage of Java reflection \cite{Arcuri-PrivateAPI}. Whilst AUTG tools such an Randoop and Evosuite are therefore able to overcome the problem of functional mocking, it also raises a new issue where the mocking of objects is "known to be susceptible to creating false positives"\cite{Arcuri-PrivateAPI} in determining faults that lie within the system under test. Arcuri et al. have shown in one of their previous studies that "46\% of generated regression tests that involved reflection or mocking failed due to false positives". They also show that 65\% of the failed tests was due to unexpected method calls on mocked objects whilst 35\% of failed tests were as a result of private methods changing and not matching the behaviour expected - similar to the "state" problem identified earlier by Riberio et al \cite{Arcuri-PrivateAPI}. Arcuri et al. however recognise the importance of the ability of AUTG tools and methods to be able to access private methods and perform proper functional mocking. The authors found that running the AUTG tools without PAFM (Private Access Functional Mocking) resulted in a total coverage of around 70.5\%. The inclusion of PAFM allowed for the increased coverage of uncovered branches by around 11.2\% which ultimately resulted in a total code coverage improvement of 3.3\% \cite{Arcuri-PrivateAPI}. In conclusion, whilst ultimately beneficial for increased coverage, Arcuri et al. reflect that more research is required for better PAFM operability and reflect that there is a "trade off between increased coverage and increased risk of false positives" which is not beneficial to the developer \cite{Arcuri-PrivateAPI}. 

The final issue within Functional Mocking that we would like to explore is the manual nature of mocking itself. Ramler et al. found that in an industrial setting where AUTG tools such as Randoop and Evosuite are used to test software, there was limited "reusable" aspects of the test configuration. The paper explored how a Human Machine Interface (HMI) with a lot of touch inputs for a factory machine could not generally be tested by the original AUTG software and how custom "adapter" programs for the GUI and touch operability had to be "mocked" in order to let the AUTG test the software. This was seen in the inclusion of an adapter program made with TestFx to send over mouse click and touch event information into Randoop as the tool itself did not support GUI Testing \cite{Ramler-AUTG-Retrospective}. This meant that a custom non-reusable solution had to be made specifically for this application by developers. Randoop itself comes with a "generic oracle in form of built-in contracts that can be used to reveal basic programming errors" - This however means that developers are required to spend more time to add in custom error oracles that match what the developers are testing the HMI against \cite{Ramler-AUTG-Retrospective}. Whilst ultimately the test was successfully applied with the custom configuration, a lot of human intervention was required and thus the amount of functional mocking and adapting that was present in the testing raises the question if the process of AUTG still retains its "automatic" nature. In real-world specific applications such as this industrial factory application, developers would be required to spend time developing custom "adapters" to allow for their software to be tested by AUTG tools whilst also needing to specify their own internal error oracles. 

In conclusion, the issue of functional mocking and the software environment remains a challenge for AUTG implementations due to the additional measures that are required as well as their limitations. More focused research is required in this space and more studies are required to see the performance of AUTG methods in a real environment with interactivity with their host environment rather than using code bases that avoid these I/O interactions. Extra research should also be conducted to allow AUTG flexibility and adaptability with different components so that they can be easily and readily tested without additional hassle. 

\subsection{Web DOM \& API}

Another challenge faced by AUTG methods that warrants its own section is the issue of interactivity with webpages and RESTful APIs. Martin-Lopez et al. states that RESTful APIS are a key element for "software reusability, integration as well as enabling new consumption models such as wearables and smart home apps" and that testing them is "critical due to their key role in software integration" \cite{Martin-Restful}. Swathi and Tiwari also build upon this sentiment by reflecting that "web applications have become crucial to most enterprises" and thus need to be rigorously tested to ensure proper functionality and expected behaviour is observed \cite{Swathi-GA-Optimise}. 

Web Application Programming Interfaces (APIs) are heavily utilised within applications today as they provide a "standard mechanism to implement create, retrieve and delete (CRUD) operations" and thus are required to be tested extensively to ensure expected behaviour with their execution \cite{Martin-Restful}. 
APIs generally contain "inter-parameter" dependencies which are "constraints that restrict the way in which two more input parameters are combined to form valid calls" \cite{Mirabella-DNN-Restful}. An example of this provided by Mirabella et al. is in the Google Maps API where if the "location" input parameter is set, then the "radius" parameter must also be set for the call to be valid. These inter-dependencies parameters are also extremely common being present in every 4 out of 5 APIs \cite{Mirabella-DNN-Restful}. The test cases for these APIs are currently generated via a black box approach where the "specifications of the API under test" is used to create test cases. This is as a result of the input parameters being specified within an "OpenAPI Specification" (OAS) document \cite{Mirabella-DNN-Restful} \cite{Martin-Restful}. Whilst black-box testing is the industry and research standard currently for APIs, Mirabella et al. have identified that black box testing does not take into account the "inter-parameter dependencies" that is required in each call as it is not specified in the OAS document. Martin-Lopez et al. also agrees with this statement and has found that the OAS can sometimes not have enough information about an API's specification \cite{Martin-Restful}. This thus means that most of the queries are actually invalid and black box approaches are currently simply brute forcing their way into generated valid test cases \cite{Mirabella-DNN-Restful}. Mirabella et al. goes on within their paper to introduce a DNN that can predict whether or not a generated API request is valid. 

Whilst black box testing is not the focus of this paper, the above context is necessary to understand the current challenges that AUTG methods face in regards to APIs. Black-box testing still remains the de-facto industry and research standard for testing APIs due to the following reasons detailed below. Martin-Lopez et al. found that the key strength of black box testing was tis ability to be applied to any API as it did not require internal access to the source code like white box testing does \cite{Martin-Restful}. Despite the limitations of black box testing as outlined by Mirabella et al. earlier, black box testing remains simple to use and doesn't suffer from an enlarged search space that white box testing AUTG methods have to endure when searching for appropriate parameters. Martin-Lopez et al. also dispels an earlier study made by Arcuri et al. to which Arcuri had found that in a comparison between black-box and white-box testing, the "latter always outperformed the former both in terms of code coverage and fault finding" \cite{Martin-Restful}. Arcuri et al.'s testing methodology was deemed to be flawed and "naive" where their Random Testing approach was simply just utilising basic random generation. Martin-Lopez et al. instead re-conducted that same testing with a more appropriate state-of-the-art tooling for both white box and white box testing and found that black box testing was significantly faster than the former \cite{Martin-Restful}. The authors state that with a low budget of calls (before being locked out by the API resource), the black box technique would achieved between 1.3\% to 55.9\% higher coverage than the white box counterpart. This was as a result of the the ability of black box testing to utilise the OAS document to tailor its approach to parameter requirements whilst white-box techniques would struggle on "complex code branches and input requirements" which was common in large code bases \cite{Martin-Restful}. Both approaches however aren't really "automatic" and still suffers from the requirements on needing similar custom "adapter" programs as raised by Ramler et al. earlier. White-box AUTG techniques thus suffer from performance issues as compared to their black box counterparts and therefore require further research before it can be utilised more effectively in web API test generation. 

Another challenge present is the inability of AUTG methods to test JavaScript functions. Fard et al. reflect that "runtime interactions between JavaScript and the DOM is error prone and challenging to test"\cite{Fard-JS-Fixtures}. The authors state that in order for the Document Object Model (DOM) of a web page to be adequately tested, that exact DOM instance would be required to be provided into the test as otherwise dynamic DOM API functions such as "getElementById" would actually return a null as the variable is not available \cite{Fard-JS-Fixtures}. Mirshokraie et al. agree with this statement and also conclude that the whole DOM is too dynamic and interactive. The existing test cases as a result are thus based around a series of event-based test cases which form an integration test - The fundamental JavaScript code remains untested at the unit level \cite{Mirshokraie-JSEFT}. Mirshokraie et al. attempt to provide a solution to this problem by providing an AUTG framework known as JSEFT (JavaScript Event and Function Testing). The authors have found JSEFT to be a robust platform achieving 100\% with no false positives which is in contrast to the false positive issues that Arcuri et al. faced earlier in private API methods. JSEFT has also been found to achieve an average code coverage of 68.4\% to which would improve with a larger search time budget \cite{Mirshokraie-JSEFT}. Fard et al. on the other hand, present a Dynamic Symbolic Execution (DSE) solution called ConFix which aims to cover all DOM-dependent paths within the JavaScript functions. Fard et al. argues that the JSEFT implementation suffers from the assumption that the DOM captured by the test at runtime would contain all of the DOM elements, values and relations \cite{Fard-JS-Fixtures}. Fard et al. argues that this may not always be the case with changing dynamic DOM insertions as well as the limitation where the entire DOM being captured could be "too large and difficult to read" \cite{Fard-JS-Fixtures}. As identified by Zhang et al. earlier however, DSE also suffers from aspects of code that cannot be symbolised or of which requires additional "infinite" memory to keep track of such as pointer variables and environmental components \cite{Zhang-Smartunit}. JavaScript in general is a dynamically-typed language which offers suffers from inherent limitations to be explored further below. Progress has been made to test DOMs but further research is necessary to best adapt AUTG methods and frameworks such as JSEFT and ConFix to obtain higher coverage. 

In conclusion, in review to the topic of RESTful APIs and web DOMs, further research is required in the API space to better improve the performance of white-box methods. White box AUTG techniques currently suffer from search space problems and the inability to create complex parameters that API expect and thus suffer in performance compared to their black box counterparts. The web DOM itself also suffers from problems where AUTG techniques are not able to directly test code. This has however been addressed with implementations such as JSEFT and Confix from researchers. The existing limitations however still present a major challenge for AUTG methods. 

\subsection{Readability}

Another dominant issue faced by AUTG is the readability of the generated unit tests. One of the main goals for AUTG was to alleviate pressure off of the developers and assist them in creating usable test cases so that they could allocate their efforts and attention to other tasks at hand. The problem however is that the test cases that were generated as a result of these algorithms were, in a sense, "gibberish" to which the developer could not ascertain their purpose and what they were doing in its execution. Whilst the test case may sometimes work as intended and find faults with the system under test, the developers are unable to understand the test case and its purpose in relation to the code which makes it hard to document what specifically the test case is doing. This ultimately results in more work for the developers as more time and effort are spent in an attempt to properly understand and document the automatically generated test cases. This goes against the goals of AUTG which is to allow the developers more time to do other things. 

Research surveys identifying the acceptance of manual and automatic testing in developer workflows have confirmed the above sentiment. Raminez et al. concludes that within the acceptance of automatic SBST results, "the time needed to understand the intentions of non-manual test cases is usually greater due to the lack of comments and descriptive names within the code" \cite{Ramirez-Interactivity}. This is caused as a result of automated tools such as Randoop and Evosuite not generating any comments to describe the generated test case code. The lack of comments inhibits the ability of developers to understand generated code which leads to more time spent attempting to understand the generated code as stated earlier. 

\begin{figure}
    \centering
    \includegraphics[width=0.7\linewidth]{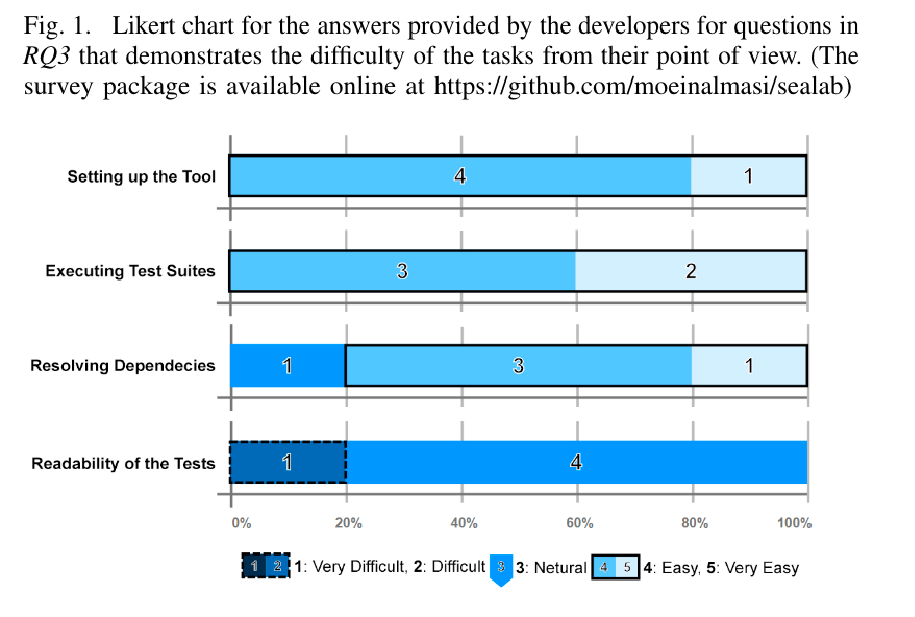}
    \caption{Survey Results on the Problems that the Developers Faced \cite{Almasi-Financial}}
    \label{fig:readability-survey}
\end{figure}

In a change from the research-applied contexts of previous papers discussed, we have found a setting where a major study was performed in an industry setting belonging to a financial application. Within the study in question, Almasi et al. deployed Randoop and Evosuite in an attempt to determine many of the 25 faults within the financial application was able to be identified and mitigated by test cases generated from the tools. Whilst Almasi et al. discusses the ability of the tools to identify "easy" and "hard faults in the earlier sections of their paper, they also conducted a survey with the developers of the financial application under test to understand their viewpoints. The authors report that the developers were happy with the process of setting up the AUTG tools for deployment by found that the readability of the generated test cases was poor \cite{Almasi-Financial}. Attached from the paper is also a chart that showed the responses of the five developers as seen in Figure~\ref{fig:readability-survey}. 

Within the survey, the developers reported that the assertions generated by the automatic tools were "poor and un-meaningful". They also reflected their dislike for the assertions that were generated and comment that they believe the assertions to be simplistic in operation. Most of the assertions were only there to check whether a list was present or whether it was a of certain size whilst other complicated assertions complied with the signature of the function under test but ultimately did not achieve anything meaningful with its assertion \cite{Almasi-Financial}. Fraser and Arcuri also comment that there is a need for "good" test cases due to the fact that in the event of a test case failing, the developers would need to step through the test case during debugging in order to find and better understand the errors present within their code \cite{Fraser-Evosuite-Challenges}. A "poor" testcase with simplistic or irrelevant assertions would not be able to assist the developer and instead fills up the code up with generated bloat.

Overall, we believe that one of the most important findings reported by Almasi et al. was that when asked if the developers would keep the automatically generated test cases within their financial application, the developer responded with a resounding "no". The developers concluded that "generated tests are not as good as manually-written tests in terms of test data" with some of the generated test cases not achieving anything meaningful with its assertions. The developers also found the generated test cases hard to read which relates back to the original problem of poor readability due to the lack of comments within the generated code. The developers were also spending additional time understanding the the generated test cases which ultimately contravenes the advantages that AUTG tools offer compared to manually creating those test cases. 

We have identified that the sample size of this survey is small with its composition limited to the five developers involved in the development of the financial application. As such, we conclude that this survey does not contain the necessary additional insights from other members of the community to form a whole cohesive view. Whilst nevertheless limited in sample data and weight, this paper is one of the few papers that have attempted AUTG within an industrial application and not a research-based context. We therefore believe in our opinion, the issues identified and raised by the developers form valid and interesting insights that are very reflective and applicable to the fundamental limitations of AUTG that this section is presenting.

\subsection{Static VS Dynamically Typed Languages}

Another limitation that AUTG posses is its ability to accurately generated tests for languages that are dynamically-typed with changing functional parameters. AUTG tools such as Evosuite and Randoop have been developed to use Search-Based Testing and Random Testing respectively (alongside other algorithms) and expect explicit type declarations within the function that they are being called to test. This is so that they can generate the appropriate parameters of a type that is expected by the calling function. Another limitation that we will explore however is that these two predominately used AUTG tools are only applicable within a Java programming language context which is a statically-typed language. There has been however, a growing popularity in using other languages which are dynamically-typed such as JavaScript and Python. 

So why does this represent a problem? If the developer were to stick with a Java environment or utilise another statically-typed language with explicit type declarations for each function that they would like to generate test cases for then everything would be fine. This is however not the case. As of July 2020, the IEEE Spectrum Ranking has ranked Python as the most popular programming language with its usage set to grow in the future through its widespread application in Machine Learning and Data Analysis \cite{Lukasczyk-Python-Automated}. This indicates the increasing usage of dynamically-typed languages to which simply depending on status-quo with Evosuite and Randoop alone would not be enough. Lukasczyk et al. provides a solution to this problem through their development of Pynguin which is an "automated test generation framework" similar to Randoop and Evosuite but for the "dynamic" programming language Python. As explored earlier in a performance comparison between search-based and random testing approaches, Lukascyzk et al. found that if the code did not possess good type information, then the search method would just be as bad as a random approach \cite{Lukasczyk-Python-Automated}.

Returning back to the point at hand, the issue of type declarations is important as it directly affects the optimisation and the "effectiveness" of the tests generated. In their paper introducing Pynguin, Lukascyzk et al. reports that within AUTG tools such as Evosuite, Randoop and now Pynguin, type information is important as it is used for the selection of parameters for function calls as well as for in the generation of complex objects \cite{Lukasczyk-Python-Automated}. The authors also further state that the absence of this type information would result in a generator effectively without guidance and would be the guessing the requirements of the function. The tools would be randomly selecting calls in a random attempt to generate the correct parameters for the function call \cite{Lukasczyk-Python-Automated}. In another paper by the same authors discussing the Pynguin AUTG tool, they reflect that this problem is a hard problem to solve as AUTG on dynamically-typed languages "aren't type-declared and can change at various stages such as in runtime and in execution" \cite{Lukasczyk-Pynguin}. This thus results in further manual intervention that is required of the developer in order to mitigate these issues and allow the testing process to run. 

\begin{figure}
    \centering
    \includegraphics[width=1\linewidth]{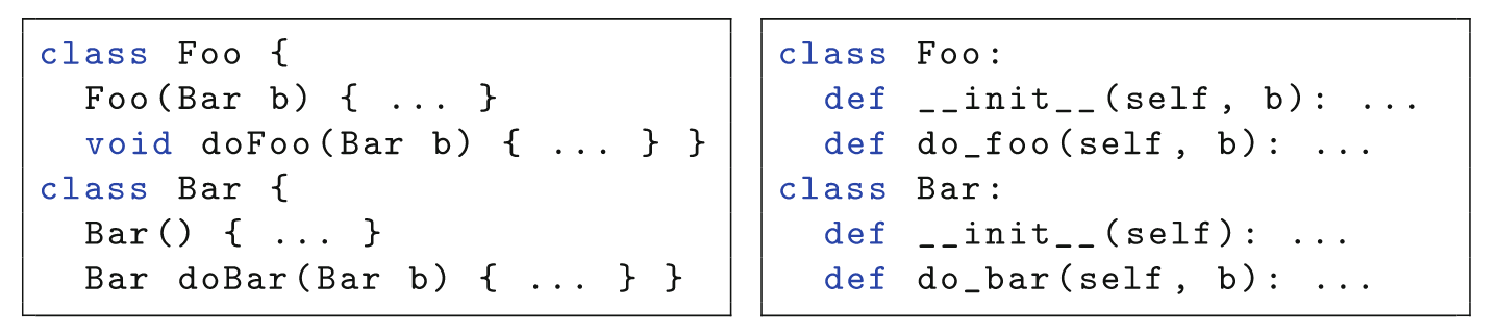}
    \caption{Typecasted Vs Non-Typecasted Functions \cite{Lukasczyk-Pynguin}}
    \label{fig:Foo-Typecast}
\end{figure}

Within the following code example of the two functions seen in Figure~\ref{fig:Foo-Typecast}, the left function code represents Java code that has type-declared functions and object instantiation. This means that each of the function parameters that are called all have descriptors of what type of object that the function is expecing. Line 3 shows that the "doFoo" function expects a "Bar" object to be passed in. Within the same figure in contrast, the Python code lacks the type declarations. Lukascyzk et al. conveys the importance of type declarations based on the two possible orientations of code generation. Within a Random Testing-based scenario such as the usage of Randoop, the "forward way" of generation would see that the only line that can be called within the code would be the "Bar()" function call seen in line 5. This is due to all of the other function calls requiring parameters to which the Random Testing algorithm knows it does not possess. With an instance of Bar initialised, the algorithm can now choose to execute the "doBar" function with a Bar object as a parameter or choose to initialise an object of "Foo" or even another object initialisation of "Bar". As per the random nature of the Random Testing method, the algorithm can continue on with calls by incorporating objects it has already made or by starting the entire process again with a fresh initialisation of a "Bar" object. Alternatively, evolutionary testing methods and tools such as Evosuite on the other hand could potentially start a call with "doBar" or "doFoo". In both cases, the evolutionary process executes in a "backwards" fashion and understands that the methods being called require a dependency of a "Foo" or "Bar" object in order to satisfy the parameter requirements of the original function call. The call to the constructor of the calss would thus be called and prepended to the original function call so that the dependencies are met and satisfied prior to execution. 

At the conclusion of this explanation, Lukascyzk et al. emphasises the importance of type information which has allowed Random Testing and Search-Based Testing to correctly execute the program on the left seen in Figure~\ref{fig:Foo-Typecast} and to continue on generating test cases. They state that without this type information for Python, the "forwards way" and "backwards way" process of generation may result in the selection of a wrong function call. This thus indicates that all of the objects and methods within the function scope would need to be considered which may very well lead to a very large search space for the AUTG method \cite{Lukasczyk-Pynguin}. If the type information is not present as in the right code example, the algorithm would only be able to "guess which calls to create new objects, or which existing objects to select as parameters for new function calls" \cite{Lukasczyk-Pynguin}. Pynguin would also have a "very low probability of choosing the correct call" for execution. As explored in the earlier section of the performance of AUTG methods, we can highlight the importance of search spaces. A large search space and the random nature of calls being "guessed" would result in a test case suite that is unable to detect many faults within the program. Generating these test suites would also result in a large amount of time being utilised as more calls would need to be called until the right one is called. Program memory space usage would also increase to keep track of all of the potentially redundant objects created. 

Thanks to the efforts of Lukascyzk and their contribution of Pynguin, the researchers have highlighted the importance of the limitations of the lack of type information within systems under test. Whilst not an issue for statically-typed languages and tools such as Randoop and Evosuite, the issue still remains a limitation of growing concern due to the increasing usage of Python and other dynamically-typed languages as identified earlier. Lukascyzk also concludes within another paper on this topic that "dynamic systems and dynamically-typed languages can be a source of reduced productivity, code usability and code quality" which is detrimental to the advantages that AUTG has afforded the developer \cite{Lukasczyk-Dynamically-Typed}.

In conclusion, this section highlights four interesting issues that are real problems AUTG as an entire field are currently facing. The above analysis answers our third and final research question we set out to explore the current challenges that AUTG faces in general. Our analysis thus showcases the issues of environmental aspects of the software as well as elements that are required to be mocked away in order for the software under test to operate properly. We also explore web problems related to the DOM and API that are also faced as an extension. Readability problems are form the main prohibitive reasoning why developers are hesitant to transition to AUTG-based tooling whilst the final reason of dynamically-typed languages also presents a barrier. These issues thus form the main challenges and answer to our research question and it is evident that further research is required in these areas to enable future adoption. 

\section{Conclusion} \label{chap:conclusion}

\subsection{Threats to Validity}

Within this Systematic Literature Review, we do not believe that there is a threat to the external validity of our paper due to the all-encompassing nature of our keyword selections as seen in Table~\ref{table: keywords} and our selection process for selecting papers as described earlier within our Methodology. The paper retrieval would have obtained all papers that were relevant to "unit testing" in some shape or form. However, we understand that we may have missed out on certain papers as a result of our exclusion criteria. Our exclusion criteria filtered out papers that were not written in English or were written prior to 1990. As the Genetic Algorithm itself was introduced in the 1970s there is the possibility that certain papers discussing its progress may have been missed. Within this paper, we had also set within our scope and exclusion criteria that only papers that generated test cases as a result of code input were to be considered. This means that other forms of generations such as Natural Language Processing were omitted in which important insights could have been gained. 

Apart from the above external validity issues, we believe that we have an internal validity issue that demands more attention. During the filtering and selection process, the initial paper filtering from 5500 papers down to the 167 papers as outlined in our PRISMA diagram in Figure~\ref{fig:Prisma} was performed by two researchers. This ensured minimal selection bias as conflicts were generally accepted. However, the next filtering step of consumption where the final 63 papers were chosen was performed entirely by 1 researcher only and was thus subject to that particular researcher's interpretation and bias. We therefore concede that there could potentially be a selection bias and that another group of researchers following our methodology process may not end up with the final papers that we have chosen to include within this review. 

\subsection{Discussion}

Within this Systematic Literature Review, we have set out to explore the field of Automatic Unit Test Case Generation in more depth and particular the Search-Based Software Testing evolutionary methods known as Genetic Algorithm and the Particle Swarm Optimisation. Within our review, we have identified previous literature reviews that explored GA and PSO in some depth. Various papers from previous authors such as Ali et al. however have determined that more research is necessary and that there is a research gap. We have also identified the same in our research and have found that papers have not made further performance and benefit comparisons between the various AUTG techniques. There was also a lack of comprehensive comparison in the improvements and limitations that exist within this field. Although some improvements such as the usage of Neural Networks was previously explored, previous reviews made by other researchers failed to explore more of the improvements that have been made to GA and PSO. These improvements as identified by our second research question included the hybrid combination improvements of GA and PSO working together as well as the combination of Mutation Testing with GA or PSO. The primary purpose of this systematic literature review was to address that research gap by providing that analysis alongside additional contextual information for the reader. It was hoped that our research questions and analysis would improve the reader's understanding of this topic at hand. This paper also explored other improvements that attempt to resolve the contended issues of "local minima" and "genetic drift" that GA possessed. In addition to this, our paper also explored the currently popular AUTG methods GA and PSO and their methods of operation as well as performance which answered our first research question. Finally, this paper also addressed our final research question which explored the various challenges and limitations currently faced by AUTG. Our review found multiple issues pertaining to environmental variables as well as the mocking of external I/O operations for network and database interoperability. Other challenges included readability which was preventing developers from adopting automated testing due to poor readability and more work required to understand the generated test cases. 

We once again believe that our Systematic Literature Review was required due to the lack of review into the topics identified by other researchers and believe that we are targeting an information gap. We would also like to reiterate the earlier concerns made by previous researchers in which that more research is required to set a common benchmark of comparison. This was a prevalent view in the review presented by Ali et al alongside the respective SBST competitions where tools such as Evosuite were tested. Reviewing our paper, we have been able to conclude that automated testing is indeed better than random and manual approaches. A key example that we would like to re-raise again is the "Tri-Typ" triangle benchmark. Mann et al. found that Random Testing would take around 94 seconds to generate close to 5 million test cases in order to achieve 100\% path coverage. GA on the other hand other took around 2 seconds to generate 39 test cases to achieve the same coverage and thus showcases the advantage that SBST has over manual and random testing. Some researchers however have also found that the evolutionary approaches were subject to random "genetic operations" which would result in an unsatisfactory result. There thus exists a belief that poor operations results in an increased search space which would in effect cause the algorithm to take a longer period of time to arrive at a satisfactory result. Many papers thus advocated for the combination of various approaches together to create a more effective algorithm. This was effective as multiple weaknesses that a particular algorithm had were resolved with the inclusion of another approach. Some researchers such as Jatana et al. and Mishira et al. have also identified that GA has a "slow convergence" issue which would make it likely for a solution to be stuck in the local minimum. Within our analysis of the "Genetic Drift" issue, we conclude that it is inconclusive to draw a result due to a wide range of papers both arguing for and against. We thus believe that further research into this space is required. 

Apart from these improvements however, we also believe that further research is necessary into the limitations and challenges as part of the future direction. As we have discussed earlier, AUTG techniques face a plethora of limitations that span from type casting, mocking issues as well as standard readability and user fatigue issues. We personally believe that in order for AUTG to gain further momentum in the developer space, further research at reducing the "set-up" pain and more reusable components are required for mass adoption. AUTG techniques currently face issues related to readability as well as its inability to be used in certain environments. Further research is thus necessary to build upon prior research and this paper. 

\subsection{Future Work}

Within this paper, we have focused solely on a few automated testing techniques which were the Genetic Algorithm as well as the Particle Swarm Optimisation. There however exists a wide plethora of other algorithms and approaches that we had not considered due to our limited scope. These algorithms include Model-Based Testing, Requirements-Based Testing and more. There exists the possibility that analysis into those approaches may improve our understanding and resolve some of the issues that we had identified above. Apart from this however, it is our belief that more research needs to be done to tackle the "genetic drift" problem as well as the limitations and challenges identified above in order for SBST to be mass-adopted. 

In conclusion, this paper aims to increase the viewer's understanding on a present research gap within the field of automated testing. We hope that the paper has been able to give the reader a better understanding and insight on the performance improvements as well as the current issues and challenges that still remain. We would also like to thank the reader for staying with us to the end of the paper. 


\bibliographystyle{acm}  
\bibliography{references-proper}  

\end{document}